\documentclass[titlepage]{article}
\usepackage{graphicx} 
\usepackage{booktabs}
\usepackage{longtable}
\usepackage{caption}
\usepackage{url}
\usepackage{pdflscape}
\usepackage{float}
\usepackage[authoryear,round]{natbib}
\usepackage{amsmath}
\usepackage{xcolor}
\usepackage[margin=1in]{geometry}
\usepackage{multirow}
\usepackage{authblk}
\usepackage{nicefrac}
\usepackage{doi}

\title{Demographics of Mesoscale Eddies in an Eddy-Permitting Ocean Model and Reanalysis}
\author[]{Benjamin Lombardi \footnote{Corresponding author: Benjamin Lombardi, benjamin.lombardi@colorado.edu}}
\author[]{Ian Grooms}
\author[]{William Kleiber}
\affil[]{Department of Applied Mathematics, University of Colorado Boulder}

\begin{document}

\maketitle
\begin{abstract}
Ocean mesoscale eddies can be thought of as the “weather” of the ocean and strongly influence the ocean’s physics, chemistry, and biology; they influence other components of the Earth system via air-sea and sea-ice interactions, and are crucial drivers of marine heat waves.
Thus, proper modeling of eddies in both historical and future climates is crucial to accurately capturing the Earth system.
Climate projections using global coupled models with eddying ocean components are only recently starting to be more widely used. 
Despite their critical role in understanding and forecasting climate characteristics, these so-called eddy-permitting models have not been explored to verify that resolved eddies are realistic, and thus any downstream scientific testing of hypotheses in biogeochemistry, ocean physics or other associated Earth systems impacted by eddies hinge on this critical assumption. 
This paper compares observed eddies with lifetimes longer than 6 weeks present in $1/4^\circ$ satellite altimetry data with observed eddies in $1/4^\circ$ reanalysis data and ocean model output.

When compared to eddies observed in satellite altimetry data, eddies in reanalysis data and ocean model output are missing almost $30\%$ of the number of eddy trajectories. 
In addition to missing eddy trajectories, the characteristics of eddies in reanalysis data and ocean model output differ from eddies observed in satellite altimetry data. 
At a high level, eddies in reanalysis data and ocean model output tend to live longer, are larger, and are weaker than eddies in observed altimetry data. 
This paper presents a variety of statistics describing these differences both spatially and in global aggregate. 
\end{abstract}

\section{Introduction}
Ocean mesoscale eddies are ubiquitous features of the global ocean that strongly influence the ocean’s physics, chemistry, biology, and climate.
Typically around $100$ km wide, mesoscale ocean eddy lifetimes range from a handful of days up to several years.
Mesoscale eddies are a critical part of the climate system and play a large role in regulating the exchange of heat and carbon with the atmosphere \citep{Guo2024TheRO}. 
Further, they also play a significant role in the redistribution of heat, salt, carbon, and nutrients around the ocean \citep{bian,dong2014global,he2024thermal,mo2024global,cornec2021impact}. 
Due to their role in transporting nutrients, they also play a part in structuring open ocean ecosystems and influence the entire food web from phytoplankton to even large ocean predators such as sharks \citep{Mcgillicuddy16,comm,braun2019mesoscale,arostegui2022anticyclonic}. 
Thus, proper modeling of eddies in both historical and future climates is crucial to accurately capturing the Earth system.

The development in the last few decades of satellite altimetry maps has allowed eddies with a sea surface height (SSH) signature to be observed at a global scale.
Eddies are mainly generated either from current instabilities or the interaction of currents with wind and topography \citep{meta3.1}.
As such mesoscale eddies often appear as anomalies within a SSH field where anticyclonic eddies are associated with a positive anomaly and cyclonic eddies are associated with a negative anomaly \citep{meta3.1}. 
Cyclonic eddies rotate counterclockwise in the northern hemisphere and clockwise in the southern hemisphere.
Anticyclonic eddies do the opposite. While subject to seasonal and regional variability, cyclonic eddies are generally associated with  cold eddy cores, while anticyclonic eddies are generally associated with warm eddy cores \citep{ni2023generation}. 

The first global eddy database distilled from satellite altimetry was introduced by \cite{chelton}. 
This dataset initially covered 1993-2008 and was updated regularly until 2016.
The production of this database was then taken over by the Collecte Localisation Satellites/Centre national d'études spatiales (CLS/CNES) team in 2017 who released an updated version of the dataset, the Mesoscale Eddy Trajectory Atlas (now on version META3.2 \citep{meta3.2}), through AVISO. 
Both of the previously mentioned datasets detect eddies by finding positive and negative anomalies in SSH fields.
However, other studies use alternative detection methods.
Some studies detect eddies by tracking the rotation of coherent structures to separate eddies from the background by using the Okubo-Weiss parameter or rotational speed \citep{isern}.
Other studies adopt a Lagrangian approach to detect mesoscale eddies \citep{beron2008oceanic,abernathey}. There are also other methods based on vector geometry \citep{nencioli2010vector} and winding angle \citep{chaigneau2008mesoscale}. 



Global coupled models with eddy-permitting ocean components are only recently starting to be more widely used in climate projections \citep{HaarsmaEtAl16}.
Despite their critical role in understanding and forecasting climate characteristics, these eddy-permitting models have not been investigated to verify that resolved eddies are realistic.
Thus any downstream scientific testing of hypotheses in biogeochemistry, ocean physics or other associated Earth systems hinges on this critical assumption.

This paper investigates how realistic resolved eddies are in an eddy-permitting $1/4^\circ$ reanalysis and a climate simulation using an eddy-permitting ocean model.
We apply the tracking algorithm developed for the META datasets to updated altimetry maps (vDT2024), reanalysis data (ORAS5), and $1/4^\circ$ ocean model output (GFDL-OM4).
The paper is organized as follows. 
Section \ref{sec:data} describes the data and tracking algorithm used. 
Section \ref{sec:glob_stats} discusses coarse global level statistics for the eddy trajectory and attribute datasets that we extract from these SSH datasets.
Section \ref{sec:local_stats} presents spatial statistics for both individual daily eddy detections and for eddy trajectories. 
Lastly, our findings are summarized in Section \ref{sec:conc}.

\section{Materials and Methods} \label{sec:data}

\subsection{py-eddy-tracker (PET)} \label{sec:pet}
    We use the {\tt py-eddy-tracker} algorithm (PET) \citep{PET} to detect and track mesoscale eddies through their SSH signature. 
    This is the detection algorithm used in the latest version of the META dataset.
    As mentioned in the previous section, the production of the dataset described by \cite{chelton} was taken over by the CLS/CNES team in 2017 who released an updated version of the dataset, the Mesoscale Eddy Trajectory Atlas (META1.0exp), through AVISO.
    In 2018, an updated version of the dataset was released (META2.0) which featured an improved tracking scheme.
    Both versions of this dataset used spatially filtered sea level anomaly (SLA) maps. 
    In 2022, META3.1exp DT \citep{meta3.1} and META3.2 DT \citep{meta3.2} were released and distributed on AVISO+.
    Version 3.1 featured an updated detection and tracking algorithm along with the switch from SLA maps to absolute dynamic topography (ADT) maps.
    Version 3.2 features the same detection and tracking algorithm, but uses updated altimetry maps (vDT2021).
    To compile our datasets for this paper, we use the same detection and tracking criteria used in META 3.2, which is described briefly below.

    Throughout, we use the term `daily eddy detection' (DED) to indicate an eddy identified at a single time, and the term `eddy trajectory' to indicate a sequence of DEDs that are determined to be the same eddy evolving over time.
    The detection and tracking algorithm first finds DEDs in SSH data, then strings DEDs together into trajectories.
    
    The detection and tracking algorithm first applies a 2D Bessel-windowed Lanczos high-pass filter with a half-power cutoff wavelength of 700 km to the SSH field.
    DEDs are then made by locating extrema in the filtered SSH field and forming closed contours around the extrema with a $2$ mm step-size. A minimum amplitude threshold of $0.4$ cm is enforced during detection. This threshold was decreased from the $1$ cm step used in \cite{chelton} as \cite{faghmous2013} demonstrated that this threshold led to the underestimation of eddies' properties. The $0.4$ cm amplitude threshold implies a minimum of three closed contours to be considered an eddy. \cite{meta3.1} deems this as reasonable compromise between computation time and the size of an eddy when compared to the resolution and noise of the altimetry maps.
    Further, a `shape error' test is performed on the outermost contour.
    This shape error test confirms that the ratio between the areal sum deviations of the contour from its best fit circle and the area of this best fit circle is below $70\%$.
    This test aims to prevent detections of eddies with shapes too different from circles where rotation is not possible. 
    Once a detection is made, PET estimates the center of an eddy, its amplitude, its effective contour (largest closed contour of the detected eddy), and various quantities relating to its speed contour (contour of maximum circum-average geostrophic speed for the detected eddy). 
    If the geostrophic velocities are not provided with the SSH field, PET estimates the geostrophic velocities.

\begin{figure}[htbp]
    \centering
    \includegraphics[width=0.55\linewidth]{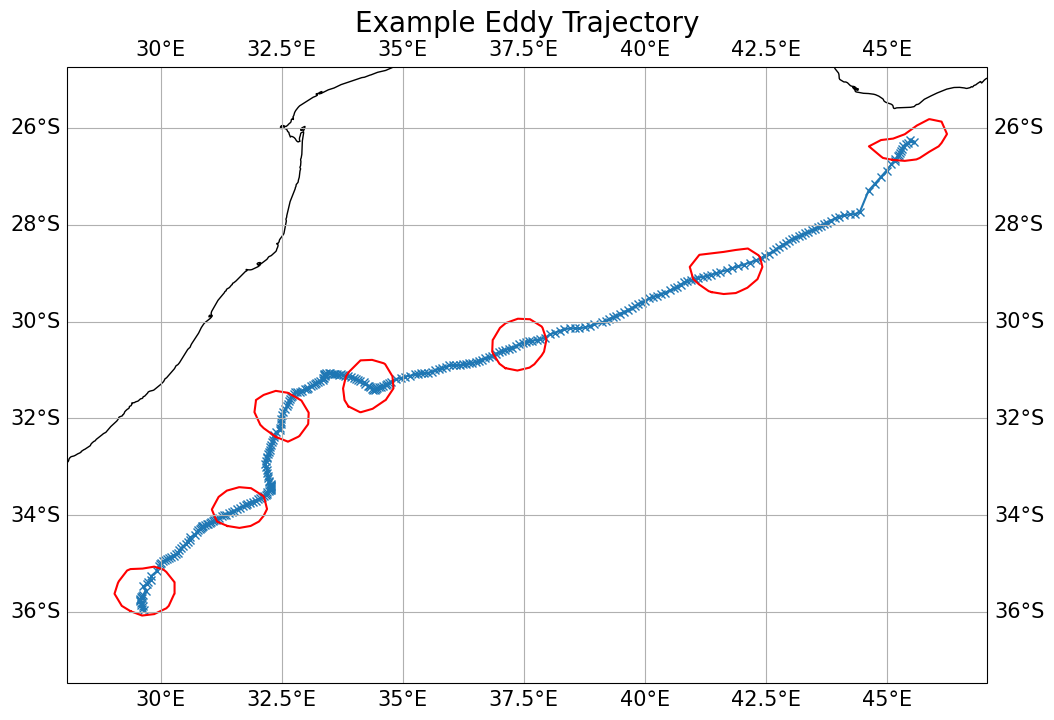}
    \caption{A cyclonic eddy trajectory with associated DEDs. Each DED is denoted with an `x'. The speed contour of the eddy is plotted every 50 days in red. The eddy trajectory travels northeast to southwest. \label{fig:ded_v_traj}}
\end{figure}

    The foregoing describes the first stage of the algorithm which identifies individual eddies (DEDs) in the daily SSH data.
    The tracking procedure used by PET to associate one DED with another as part of a trajectory is based on the overlap of the effective contours (outermost closed contour containing the SSH extrema) of eddy identifications separated by one day.
    If the intersection over the union (IOU) of the two contours is greater than $5\%$ then the observation is retained as a candidate for the trajectory. 
    If there is more than one candidate for the eddy trajectory then the candidate with the largest IOU is retained. 
    This addition of candidates to the trajectory proceeds sequentially in time; once there are no more candidates to be added, they are designated as an eddy trajectory. 
    An example eddy trajectory and component DEDs identified by PET are displayed in Fig.\ \ref{fig:ded_v_traj}. 

    To compile our data, we use the same detection and tracking parameters used in the META 3.2 \citep{meta3.2}. 
    For more details on the tracking and detection algorithm see \citep{meta3.1}.

\subsection{SSH Fields} \label{sec:ssh}
    In this paper we ran PET on 4 different SSH fields: observational estimates of (i) ADT and (ii) SLA from the Sea Level DT2024 dataset \citep{vDT2024}, (iii) the zos field (section H7.3 of \cite{griffies2016omip}) from the ORAS5 reanalysis \citep{reanalysis}, and (iv) the GFDL-OM4 ocean model \citep{gfdl-OM4}. 
    Both ADT and SLA are derived variables acquired by merging data from two different altimetry satellites. 
    These observational variables are produced by the Data Unification and Altimeter Combination System (DUACS) and includes data spanning several different altimetry missions.
    To ensure long-term stability, a satellite constellation with a stable number of altimeters is used.
    The final gridded merged product is derived from along-track altimeter measurements, which then go through several processing steps, such as removing biases and filtering.
    An optimal interpolation method is then used to map these processed along-track measurements to a regular $1/4^\circ$ latitude-longitude grid. 
    At the time of access, this dataset begins on January 1, 1993 and ends of December 31, 2023. 
    The ADT and SLA datasets were historically distributed through AVISO but can now be found on the Copernicus Climate Change Service Climate Data Store \citep{vDT2024}. 
    Details on the specific satellite missions and processing steps can be found in the Sea Level version DT2024: Product User Guide and Specification file in \citep{vDT2024}.

    It should also be noted that there are limitations in the ability of the DUACS data sets to accurately represent eddying behavior. 
    While DUACS provides maps on a global $1/4^\circ \times 1/4^\circ$ and daily grid framework, the effective resolution of the data is not necessarily equal to the $1/4^\circ$ spatial and daily temporal resolution of the data set due to the filtering that occurs during optimal interpolation.  
    The effective spatial and temporal resolution of these maps is estimated by \cite{ballarotta2025} based on the ratio between the spectral content of the mapping error and the spectral content of independent true signals. 
    The full details of this approach are described by \cite{ballarotta2019resolutions}. 
    It is estimated that the average temporal resolution is approximately 33 days and that the average spatial resolution (expressed as a wavelength) is approximately 500 km at the equator, 200 km at midlatitudes, and 100 km at high latitudes \citep{ballarotta2025}.
    To address concerns regarding the temporal resolution of these maps, we will only consider eddies that live at least 6 weeks as this the upper range of the effective temporal resolution for the altimetry maps \citep{ballarotta2025}.
    
    \cite{ballarotta2019resolutions} estimate that the smallest reliably detectable eddy has a radius 20 -- 25\% of the effective resolution, which means that eddies with radii larger than $\approx25$ km are properly resolved at high latitudes, larger than $\approx 50$ km are properly resolved in the midlatitudes, and larger than $\approx 125$ km are properly resolved near the equator \citep{ballarotta2019resolutions}. 
    There is also work that indicates that satellite altimetry data overestimates eddy size: 
    \cite{bashmachnikov2020eddies} investigate eddy dynamics in the North Greenland Sea and the Fram Strait using satellite altimetry data (AVISO), satellite aperture radar data (SAR), and a high-resolution Finite Element Sea ice-Ocean Model (FESOM). 
    They find eddies detected in AVISO on average have radii 1.5--2 times larger than radii estimated from in-situ observations in their region of interest.
    They also find that smaller scale eddies with a radii of approximately 15-20 km are robustly detected in along-track altimetry measurements, but the smoothing and interpolation process that creates the final gridded AVISO product erases these small scale structures unless the eddy has a very strong SSH signal.
    
    To avoid comparing model and reanalysis output to observations that are unreliable, we remove eddies in the middle equatorial band ($15^\circ$S-$15^\circ$N)  due to DUACS's low resolution in this region.
    We do not impose any additional constraints on the area of eddies other than what is imposed by PET (see previous section).
    Together with the restriction to trajectories at least 6 weeks long, we believe that this minimizes the risk of spurious or under-resolved eddy detections in altimetry data and helps provide a fair comparison between observations and model output.
    
    While ADT and SLA are both measures of SSH, they differ in their physical meanings. 
    Altimetry data provides the sea surface height above the reference ellipsoid.
    SLA is defined as the variable part of sea surface height around its time-mean state.
    Thus, the value of SLA depends on the reference period of the dataset.
    ADT is the height of the sea surface above the geoid, which is the shape that the ocean surface would take if it was only under the influence of the gravity of Earth.
    Mean Dynamic Topography (MDT) is the mean state of SSH above the geoid.
    The relationship between ADT and SLA is that ADT is equal to the sum of SLA and MDT.
    Although ADT and SLA are physical quantities with existence independent of the observational products that estimate them, we will use the acronymns `ADT' and `SLA' as a shorthand throughout to refer to the observational estimates described above.
    Most of our analysis is performed using ADT rather than SLA, following \cite{meta3.1}; some global results for SLA are presented early on to emphasize that the differences between the model-based SSH fields (ORAS5 and GFDL-OM4) and observation-based fields are significantly greater than the differences between eddies identified in SLA and ADT.

    The ORAS5 Reanalysis product is available via the Copernicus Marine Service Information Marine Data Store as part of the CMEMS Global Ocean Ensemble Reanalysis product of $1/4^\circ$ degree resolution.
    ORAS5 was produced by ECMWF and uses NEMO3.4.1 as its model using the ORCA $1/4^\circ$ grid with 75 vertical levels, coupled to the LIM2 sea ice model.
    It is forced at the surface by ERA interim and assimilates sea surface temperature (SST), sea level anomalies (SLA), sea ice concentrations (SIC), and in situ temperature and salinity anomalies.
    It uses the NEMOVAR (3DVar) assimilation scheme with a 5-day assimilation window.
    Similar to ADT and SLA, this dataset begins on January 1, 1993 and ends December 31, 2023. The ORAS5 data is provided on a regular latitude-longitude $1/4^\circ$ grid.
    The sea surface height measurement in ORAS5 is the zos variable described in section H.7 of \cite{griffies2016omip}. 
    zos is the effective sea surface height above the geoid as if sea ice (and snow) at a grid cell were converted to liquid seawater.
    The difference between ADT and zos is that when deriving ADT an approximation is used for Earth's geoid while the model's geoid, used to compute zos, is known. 
    More details on ORAS5 can be found in the Product User Manual for \cite{reanalysis}. 
    
    Lastly, daily zos data from a run of the GFDL-OM4 model, which couples the Modular Ocean Model (version 6) to the Sea Ice Simulator (version 2) was kindly provided by A. Adcroft.
    GFDL-OM4 is developed by the Geophysical Fluid Dynamics Laboratory (GFDL).
    We specifically consider GFDL-OM4p25 which has a nominally $1/4^\circ$ native tripolar grid and a hybrid vertical coordinate with 75 levels. More details on the model configuration are given by \cite{gfdl-OM4}.
    Our experiment was forced using the interannual Coordinated Ocean-sea ice Reference Experiment (CORE) protocol from \cite{griffies2009elements} and \cite{danabasoglu2014north}.
    The CORE atmospheric forcing is repeated 5 times and our data comes from the fifth cycle of forcing. For GFDL-OM4, we only consider data from January 1, 1993 through the end of the model run in December 31, 2007. Note that this data does not contain any leap days. Like ORAS5, the SSH variable in GFDL-OM4 is zos.
    
    Unlike ADT, SLA, and ORAS5, the GFDL-OM4 data is natively on a $1/4^\circ$ tripolar curvilinear grid.
    While PET does support eddy detection on ``unregular" grids, such as a tripolar curvilinear grid, not all of the features are supported for this type of detection. 
    Specifically on unregular grids, neither the 2D-Bessel-windowed Lanczos filter nor the method for the estimation of geostrophic velocities are available in PET.
    (The GFDL-OM4 data set provides surface velocities, but not geostrophic surface velocities.)
    In order to ensure that the same detection and tracking algorithm is applied across all data sets we choose to re-grid the GFDL-OM4 zos data to the same $1/4^\circ$ regular latitude-longitude grid as the other data sets before applying the PET detection and tracking algorithm.
    For all points south of $82^\circ$ N a bilinear regridding was used.
    North of $82^\circ$ N a nearest-neighbors regridding was used to account for the singularities in the tripolar grid.

    We have chosen to examine GFDL-OM4 and ORAS5 for several reasons.
    GFDL-OM4 was chosen as it was the only data available to us from a published eddy-permitting model that included daily SSH data. 
    The latest CMIP archive (CMIP6 at the time of writing) has no eddy-permitting model output that includes daily SSH data.
    In contrast, there are several eddy-permitting reanalysis datasets available.
    We chose ORAS5 as it is one of the more recent reanalysis datasets.

\section{Global-aggregated Statistics} \label{sec:glob_stats}
    When comparing the eddy datasets corresponding to each of the differing SSH fields, we only consider eddies that live at least 6 weeks.
    As mentioned in the previous section, the average temporal resolution of the altimetry maps is $\approx 33$ days. As this is the average temporal resolution, we choose $6$ weeks as our minimum lifetime threshold to avoid spurious detections. 
    This leads to only considering DEDs corresponding to eddy trajectories that comprise of a minimum of 42 DEDs. 
    As mentioned in the previous section, we will also only consider eddies outside of the equatorial band $15^\circ$S - $15^\circ $N due to concerns about the spatial resolution of the altimetry maps.
    We first look at statistics at an  aggregate level: the total number of DEDs and the total number of eddy trajectories.
    As the datasets span different time lengths, to ease comparison between ADT, SLA, ORAS5, and GFDL-OM4 we provide the average number of DEDs and tracks per year for all of the datasets. These quantities are listed in Table \ref{tab:track_nums}. 
    The total counts are provided in the Supplementary Material.  

    Both of the observational datasets (ADT and SLA) have a similar number of average DEDs per year, with approximately $750,000$ cyclonic eddy DEDs per year and roughly $730,000$ anticyclonic eddy DEDs per year.
    ORAS5 and GFDL-OM4 are deficient by comparison.
    (Recall that a single eddy trajectory will contribute at least 42 DEDs to the count.)
    ORAS5 has a just under $610,000$ cyclonic DEDs per year and slightly less than $575,000$ anticyclonic DEDs per year.
    GFDL-OM4 has about $565,000$ cyclonic DEDs per year and slightly less than $590,000$ anticyclonic DEDs per year. 
    When compared to ADT, ORAS5 has $\approx 81\%$ of the number of cyclonic DEDs per year and $\approx 79\%$ of the number of anticyclonic DEDs per year.
    GFDL-OM4 has $\approx 75\%$ of the number of cyclonic DEDs per year and $\approx 81\%$ of the number of anticyclonic DEDs per year when compared to ADT.
    Both ORAS5 and GFDL-OM4 are missing at least $\approx20\%$ of the number of DEDs that they are supposed to have when compared to ADT. 

    We also also interested in the number of eddy trajectories, which we will also refer to as `tracks.' ADT has about $8,100$ cyclonic tracks per year and $7,600$ anticyclonic tracks per year.
    SLA has a similar amount with about $8,100$ cyclonic tracks per year and $7,800$ anticyclonic tracks per year. 
    ORAS5 on the other hand has about $5,900$ cyclonic tracks per year and $5,500$ anticyclonic tracks per year.
    GFDL-OM4 has about $5,200$ cyclonic tracks per year, and $4,900$ anticyclonic tracks per year. 
    When compared to ADT, ORAS5 and GFDL-OM4 only respectively have $72\%$ and $64\%$ of the number cyclonic tracks that ADT has. 
    Similarly, ORAS5 only has $72\%$ and GFDL-OM4 has $64\%$  of the number anticyclonic tracks that ADT has. 
    Both the $1/4^\circ$ reanalysis ORAS5 and the $1/4^\circ$ model GFDL-OM4 are missing about $20\%$ of the expected number of daily eddy identifications (DEDs) and $30\%$ of the expected number of eddy trajectories when compared to the observational products, with GFDL-OM4 having slightly less of both than ORAS5. 
    
\begin{table}[ht]
\centering
\begin{tabular}{lrr}
\toprule
 Metric & Cyclonic & Anticyclonic \\
\midrule
Average Daily Eddy Detections (DEDs) per Year ADT & 751,690 & 723,337 \\
Average Daily Eddy Detections (DEDs) per Year SLA & 749,779 & 735,083 \\
Average Daily Eddy Detections (DEDs) per Year ORAS5 & 609,354 & 573,036 \\
Average Daily Eddy Detections (DEDs) per Year GFDL-OM4 & 565,268 & 587,301 \\
Average Number of Tracks per Year ADT & 8,115 & 7,582 \\
Average Number of Tracks per Year SLA & 8,116 & 7,815 \\
Average Number of Tracks per Year ORAS5 & 5,871 & 5,448 \\
Average Number of Tracks per Year GFDL-OM4 & 5,175 & 4,873 \\
\bottomrule
\end{tabular}
\caption{Eddy Track Statistics.\label{tab:track_nums}}
\end{table}

    We see that ORAS5 and GFDL-OM4 have too few eddies.
    However, this begs the question: which eddies are missing? 
    Are ORAS5 and GFDL-OM4 missing long-lived eddies?
    Large eddies? 
    Circular eddies? etc. 
    Further, \emph{where} are ORAS5 and GFDL-OM4 missing eddies?
    In the following sections, we will address these questions.
    In each section we provide summary statistics about the quantity of interest as well as display the globally-aggregated empirical scaled PDFs and local maps of the mean.
    We only present results for cyclonic eddies for these quantities unless otherwise noted. 
    For the quantities where only the results for cyclonic eddies are presented, the results for anticyclonic eddies are qualitatively similar. 
    However, all results for anticyclonic eddies are available in the Supplementary Material.
    Further a table displaying all of the summary statistics for both cyclonic and anticyclonic is also available in Appendix A. 
    Lastly, in the following sections we omit the results for SLA. 
    As is presented in Table \ref{tab:track_nums}, the total number of eddies and DEDs are very similar between ADT and SLA.
    Further, eddy characteristics are also globally similar between the two. 
    Given this similarity and the fact that ADT is most analogous to the model zos variable, we will only present comparisons with ADT moving forward; global statistics for SLA are also reported in Appendix A. 

\section{Spatial Distribution of Eddy Characteristics} \label{sec:local_stats}

    Globally, both ORAS5 and GFDL-OM4 have too few eddies. In this section we investigate the spatial distribution of eddy behavior in ORAS5 and GFDL-OM4 across the globe. On an eddy attribute basis, we assess how accurately ORAS5 and GFDL-OM4 represent eddy behavior.

\subsection{DED-Level Statistics}
    We begin by looking at DED-level statistics. 
    These are statistics that are calculated using DEDs rather than trajectories. 
    In the following sections we will discuss eddy census, amplitude, equivalent radius, acircularity, and speed average. 

\subsubsection{Average Eddy Census per Year} \label{sec:census}
    We first begin by looking at the eddy census. 
    Eddy census is the number of DEDs within each $1^\circ \times 1^\circ$ longitude-latitude box.
    To make a fair comparison across the datasets, we divide each of these census values by the number of years in the dataset.
    The ADT and ORAS5 datasets span 31 years and GFDL-OM4 spans 15 years.
    
\begin{figure}[htbp]
    \centering
    \includegraphics[width=\linewidth]{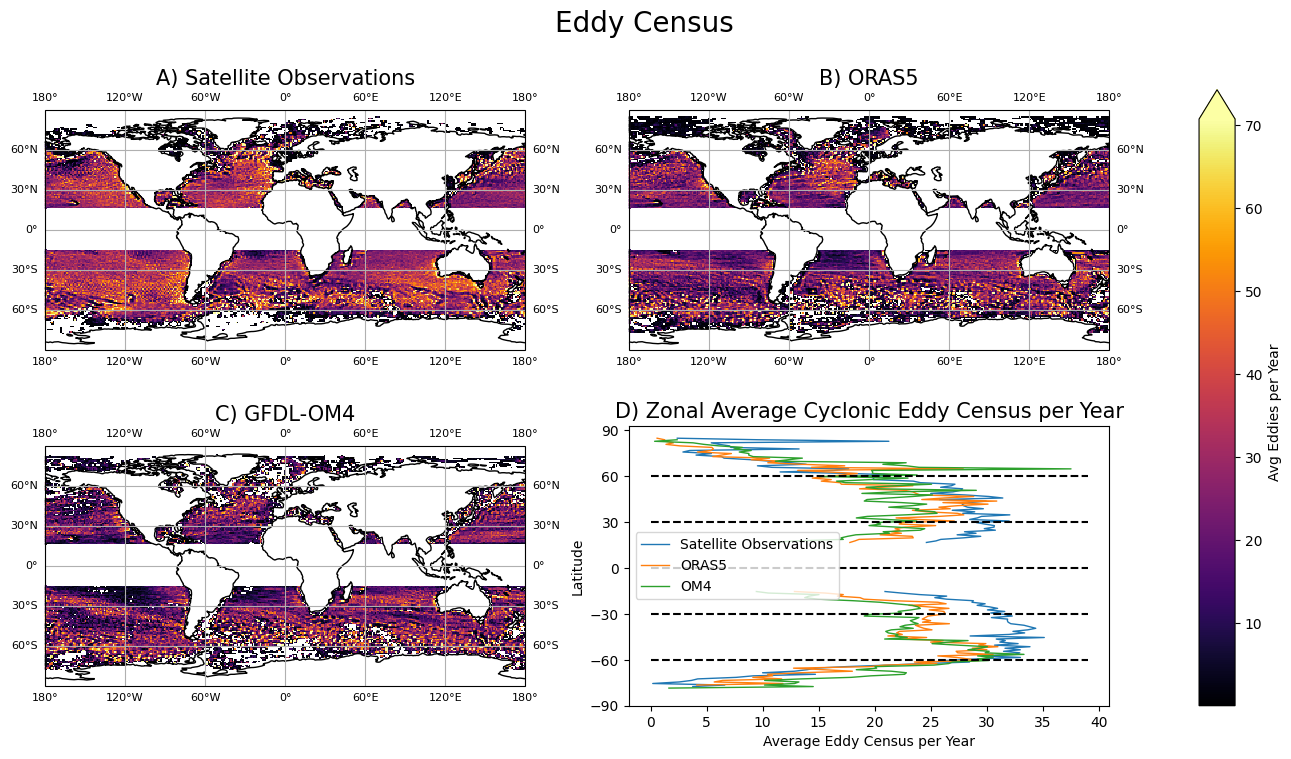}
    \caption{Cyclonic Eddy Census. (D) is the zonal average cyclonic eddy census per year. (A-C) depict for the eddy census per year within each $1^\circ \times 1^\circ$ longitude-latitude box for ADT, ORAS5, and GFDL-OM4. \label{fig:census}}
\end{figure}

    The zonal average cyclonic eddy census per year is depicted in the of Fig.\ \ref{fig:census}d. 
    ADT has more DEDs than ORAS5 and GFDL-OM4 between $60^\circ$ S and $60^\circ$ N.
    This difference is the largest in the tropics.
    At extreme latitudes outside of the $\pm60^\circ$ band all three of the datasets have a similar number of DEDs with ADT generally having slightly fewer DEDs than the other two datasets. 
    The census plots are also displayed in Fig.\ \ref{fig:census}a-c. ORAS5 and GFDL-OM4 are missing eddies almost everywhere.
    They have a similar trend of having fewer eddies at lower latitudes and then generally an increasing number of eddies until $60^\circ$ N and $60^\circ$ S, consistent with the zonal-mean plot in Fig.\ \ref{fig:census}d.
    ORAS5 has notably more eddies in the northeast Atlantic and in the Pacific subpolar gyre than GFDL-OM4.
    
    To complement the qualitative understanding gained by a visual inspection of these maps, we provide two quantitative measures of the differences.
    First we compute the anomaly cosine correlation, also sometimes known as anomaly pattern correlation, of both the ORAS5 and GFDL-OM4 census with the ADT census. As we are comparing the anomalies relative to each dataset's mean, this is equivalent to the Pearson correlation. 
    Thus, standard interpretations follow: values of correlation close to $1$ or $-1$ indicate strong linear or anti-linear relationships, respectively.  
    When computing the anomaly cosine correlation, we only consider entries where both maps are not NaN (NaN's occur where no eddies are detected) as well as weight each grid cell by its respective area.
    The census anomaly cosine correlation with ADT is $0.62$ and $0.53$ for ORAS5 and GFDL-OM4 respectively. 
    The anomaly cosine correlation of these maps is low and indicates that ORAS5 and GFDL-OM4 could not be fixed by simply shifting and scaling their existing spatial patterns of eddy density.
    This can be inferred from the zonal-mean plots, which show that while ORAS5 and GFDL-OM4 have a similar number of eddies at high latitudes, they are missing eddies almost everywhere else.
    If their spatial pattern were correct they would have approximately 70-80\% of the eddies uniformly across all latitudes, rather than being deficient at some latitudes and correct at others. 
    Proportionally to their total number of eddies, ORAS5 and GFDL-OM4 are generating too many eddies at higher latitudes than they are supposed to when compared with ADT.
    The second quantitative measure of the differences between these maps is the root mean square error (RMSE).
    RMSE complements correlation by measuring the magnitude of the differences between the maps, where correlation simply measures the differences in the spatial patterns. 
    The RMSE of these maps when compared to ADT is $15.18$ and $17.02$ DEDs per year for ORAS5 and GFDL-OM4 respectively.
    
\subsubsection{Eddy Amplitude} \label{sec:amp}
\begin{figure}
    \centering
    \includegraphics[width=\linewidth]{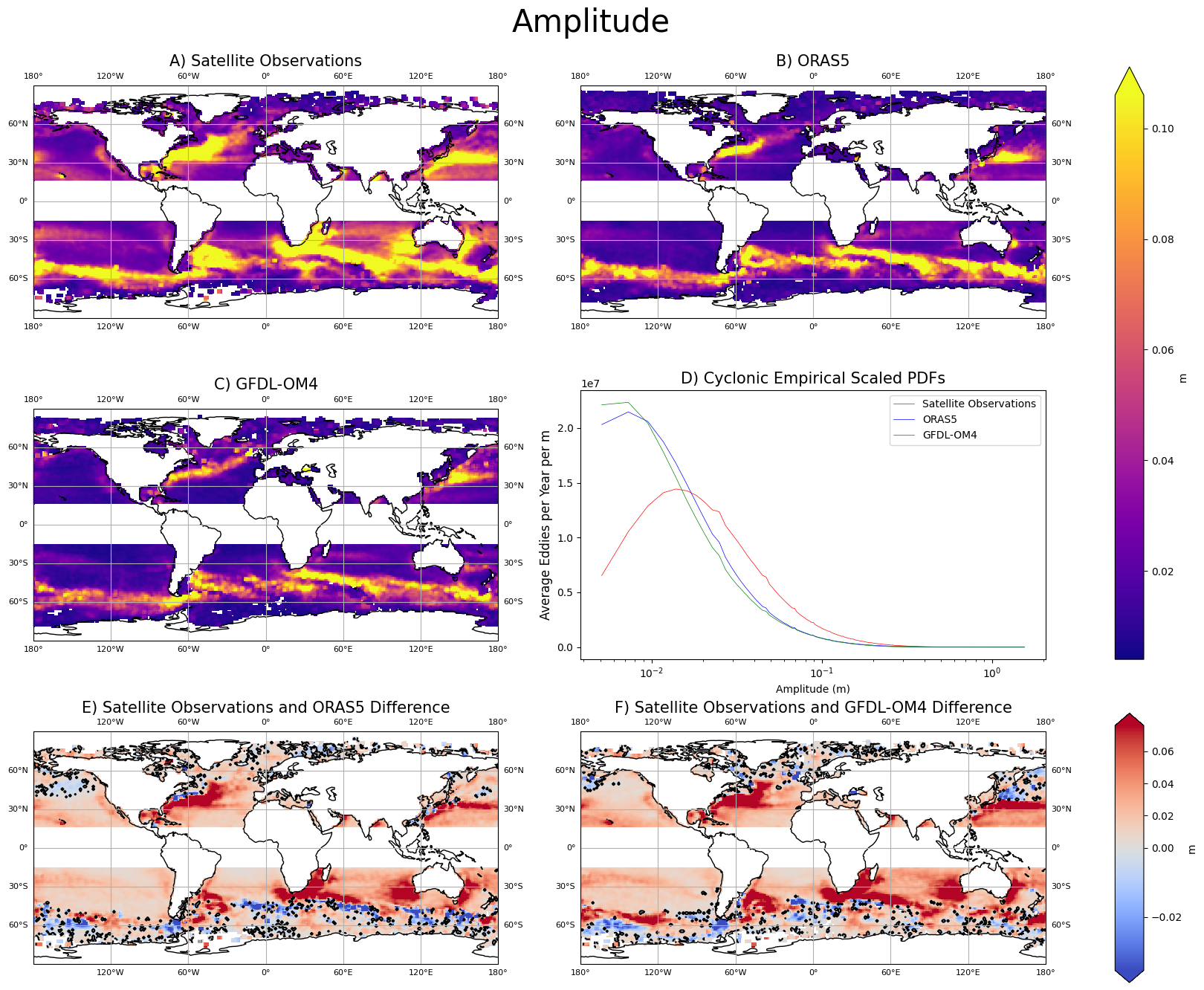}
    \caption{Cyclonic Eddy Amplitude.  (A-C) display the spatial maps of mean amplitude for ADT, ORAS5, and GFDL-OM4. (D) shows the scaled empirical global PDFs of amplitude for ADT, ORAS5, and GFDL-OM4. (E-F) show the differences between ADT and ORAS5 maps and the differences between ADT and GFDL-OM4 maps for amplitude. Areas that are grayed out do not have differences that are statistically significant.
    \label{fig:amp}}
\end{figure}

    The amplitude of an eddy is defined as the magnitude of the height difference between the extremum of SSH within the eddy and the SSH around the effective contour defining the eddy edge, in meters. 
    Fig.\ \ref{fig:amp}d shows the global empirical scaled PDFs for amplitude from ADT, ORAS5, and GFDL-OM4.
    These PDFs are scaled so that integrating the PDF between two amplitude values $a_1$ and $a_2$, where $a_1 < a_2$, yields the number of eddies per year that have an amplitude between $a_1$ and $a_2$.
    Although we have seen that ORAS5 and GFDL-OM4 both have far fewer DEDs than ADT, the scaled PDFs show that ORAS5 and GFDL-OM4 have an excess of DEDs with amplitudes less than about 1 cm.
    This is more than offset by their lack of DEDs at larger amplitudes.
    
    PET identifies a substantial number eddies with an amplitude less than 1 cm. In the satellite-derived ADT product, these DEDs only make up approximately $8.65\%$ of the total number of DEDs.
    This may seem unrealistically small for a mesoscale eddy, but a DED having an amplitude of less than 1 cm does not imply that an eddy had of amplitude of less than 1 cm  throughout its entire lifetime. 
    \cite{samelson2014randomness} finds that the amplitude of an eddy generally starts small, increases to some equilibrium value, and then decreases towards the end of its lifetime.
    In ADT, about $50.3\%$ of DEDs with an amplitude less than 1 cm occur either in the first or last $15\%$ of an eddy's lifetime. 

    We next consider global amplitude statistics that are independent of the total number of DEDs, and that instead describe the distribution of amplitudes for each population of DEDs.
    For cyclonic eddies the global amplitude means for ADT, ORAS5, and GFDL-OM4 are $6.04$ cm, $4.14$ cm, and $3.64$ cm, the medians are $3.62$ cm, $2.17$ cm, and $2.04$ cm, and the standard deviations are $7.37$ cm, $5.89$ cm, and $4.38$ cm, respectively.
    Globally, in both ORAS5 and GFDL-OM4, the amplitude of eddies is generally weaker.
    This is consistent with the incorrect shape of the PDFs shown in Fig.\ \ref{fig:amp}, where the distributions of ORAS5 and GFDL-OM4 both peak just after the smallest allowable amplitude of 4 mm.
    Not only are the global means smaller for both ORAS5 and GFDL-OM4, but the global standard deviations are smaller as well. 
    
    Globally, both ORAS5 and GFDL-OM4 have deficient amplitude. 
    To assess the spatial distribution of eddy amplitudes we take the mean amplitude of all DEDs within a $3^\circ \times 3^\circ$ longitude-latitude box. 
    We do this for overlapping $3^\circ \times 3^\circ$ boxes on a $1^\circ$ grid and plot the results in Fig.\ \ref{fig:amp}a-c.
    Fig. \ \ref{fig:amp}e-f shows the differences in local mean between ADT and ORAS5 and GFDL-OM4 respectively. 
    A positive difference indicates that ADT has a larger mean and a negative difference indicates that the mean of ORAS5 or GFDL-OM4 is larger.

    To quantify how well ORAS5 and GFDL-OM4 capture the local structure of the global amplitude distribution when compared to ADT, we compute the anomaly cosine correlation and root mean square error (RMSE) for these maps. 
    The anomaly cosine correlation of ADT with ORAS5 is $0.79$ and there is a RMSE of $2.77$ cm.
    The anomaly cosine correlation of ADT with GFDL-OM4 is $0.67$ with a RMSE of $3.41$ cm.
    The anomaly cosine correlation of ORAS5 and GFDL-OM4 are relatively good compared to the other quantities discussed later in this paper. 
    Although eddy amplitude is too low in ORAS5 and GFDL-OM4, with RMSE on the same order of magnitude as the global mean, the spatial distribution of where low and high amplitude eddies occur is relatively accurate when compared to other quantities included in the following sections.

    While ORAS5 and GFDL-OM4 both capture many of the larger scale features of the amplitude distribution, they both struggle to accurately capture the finer local distributions of eddy amplitude outside of the major current regions. 
    ORAS5 and GFDL-OM4 are missing amplitude nearly everywhere, with a few exceptions.
    GFDL-OM4 appears to have higher amplitude than it should north of the Kuroshio extension, and both ORAS5 and GFDL-OM4 have too-high amplitude in parts of the Southern Ocean.
    ORAS5 also has overly large amplitudes just north of the North Atlantic Current east of the northwest corner.
    We conjecture that these excessive amplitudes may reflect biases in the position of the associated current systems: the Kuroshio Extension, Antarctic Circumpolar Current, and North Atlantic Current.

     We test for the statistical significance of the differences of means in each box using Welch's test. This test does not assume equal variances. 
     Due to the large number of multiple tests, spuriously significant $p$-values are expected even when there are no differences between data sets. 
     To control for multiple testing, we follow the False Discovery Rate (FDR) approach \citep{benjamini1995controlling,wilks2016stippling}.
     This approach operates on a collection of $N$ hypothesis tests with $p$-values $p_i$.
     These $p$-values are then sorted in ascending order which will be denoted as $p_{(1)} \leq p_{(2)} \leq ... \leq p_{(N)}$.
     Local null hypotheses are rejected if their $p$-value is smaller than $p_{FDR}^*$ which is based on the distribution of the sorted $p$-values. 
     $p_{FDR}^*$ is calculated as 
     \begin{equation*}
         p_{FDR}^* = \max_{i=1,...,N} \left [ p_{(i)}: p_{(i)} \leq \frac{i}{N}\alpha_{FDR} \right]
     \end{equation*}
     where $\alpha_{FDR}$ is desired control level for the FDR.
     \cite{wilks2016stippling} recommends that for processes with moderate to strong spatial correlation that approximately correct results can be achieved by choosing $\alpha_{FDR} = 2\alpha_{global}$. 
     We choose a global significance level of $\alpha_{global} = 0.05$ which gives a $\alpha_{FDR} = 0.1$.
     In Fig. \ \ref{fig:amp}e-f differences that are not significant are grayed out. 
     The differences between ADT and ORAS5 and ADT and GFDL-OM4 are significant almost everywhere with some small exceptions in the ACC and at extreme northern latitudes.

\subsubsection{Eddy Equivalent Radius} \label{sec:area}
\begin{figure}[htbp]
        \centering
        \includegraphics[width=\linewidth]{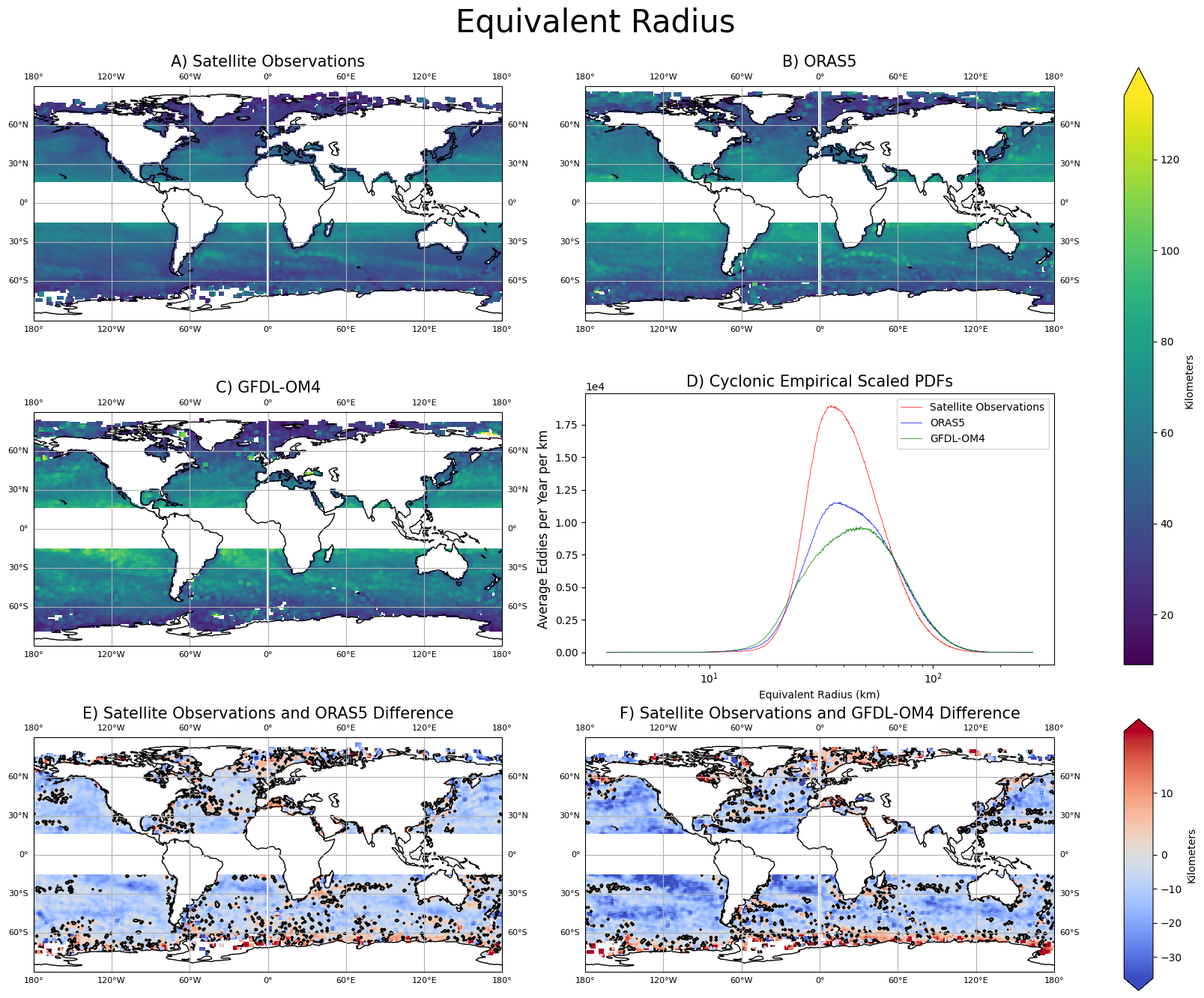}
        \caption{Cyclonic Equivalent Radius.  (A-C) display the spatial maps of mean equivalent radius for ADT, ORAS5, and GFDL-OM4. (D) shows the scaled empirical global PDFs of equivalent radius for ADT, ORAS5, and GFDL-OM4. (E-F) show the differences between ADT and ORAS5 maps and the differences between ADT and GFDL-OM4 maps for equivalent radius. Areas that are grayed out do not have differences that are statistically significant. \label{fig:area}}
    \end{figure}

    Eddy equivalent radius is defined to be the radius of a circle with the same area as the area enclosed by the speed contour of an eddy.
    The speed contour of an eddy is the contour with the highest average geostrophic velocity. 
    The global scaled PDFs in Fig.\ \ref{fig:area}d show that ORAS5 and GFDL-OM4 are mainly missing eddies with an equivalent radius  between $\approx 20$ and $\approx 70$ km.
    Despite having too few eddies overall, GFDL-OM4 has a slight excess of eddies with a radius $<20$ km, though these eddies may not be well-resolved in the observational product, as mentioned in section \ref{sec:data}.
    This is similar to how GFDL-OM4 has an excess of small-amplitude eddies, though the difference is not so severe.
    Both GFDL-OM4 and ORAS5 also have a slightly larger number of eddies with equivalent radii larger than about $70$ km.  

    We next consider the global distribution of equivalent radius for each population of eddies, independently from the total number of eddies. 
    For cyclonic eddies in ADT, ORAS5, and GFDL-OM4, the global means are 50.71 km, 57.26 km, and 58.81 km the medians are 46.14 km, 52.42 km, and 54.85 km, and the standard deviations are 20.61 km, 25.11 km, and 25.31 km, respectively.
    Unlike amplitude, eddies in ORAS5 and GFDL-OM4 are wider than they are in ADT and their radii have a higher standard deviation than ADT.

    The spatial maps for cyclonic eddies are displayed in panels of Fig.\ \ref{fig:area}a-c. The difference of these maps are displayed in panels E-F).
    Both ORAS5 and GFDL-OM4 tend to have larger eddies when compared to ADT throughout the ocean, but especially in the subtropical south Atlantic and south Pacific oceans. These differences remain statistically significant despite there not being many eddies there in the model and reanalysis, as shown in Fig.\ \ref{fig:census}.
    As was with the case with amplitude, the differences in mean in Fig. \ \ref{fig:area}e-f are statistically significant almost everywhere. 
    There are small regions where the differences are not significant. 
    However, there are no large regions where this is the case. 
    The areas where the differences are not deemed significant tend to be at higher latitudes where there are not as many eddies.
    
    

    While the mean plots portray a stark picture, the anomaly cosine correlation of these plots is relatively good. 
    Comparing ORAS5 to ADT yields a anomaly cosine correlation of $0.82$ and a RMSE of 7.37 km.
    Comparing GFDL-OM4 to ADT yields a anomaly cosine correlation of $0.73$ and a RMSE of 10.6 km. 
    This suggests that the spatial structure of eddy equivalent radius is good in both ORAS5 and GFDL-OM4, but their eddies are too wide nearly everywhere.
    
    As the DUACS dataset that ADT is derived from is not able to accurately resolve smaller eddies due to limitations in its spatial resolution, one might expect observed eddies in ADT to be wider than those observed in the model and that assimilating altimetry data in ORAS5 might increase eddy size. 
    However, the opposite of this appears to be the case. 
    Eddies in ADT are smaller than those detected in both ORAS5 and GFDL-OM4. Further, assimilating altimetry data appears to reduce eddy mean width by adding small eddies. As we expect eddies observed in ADT to be wider than they are in reality, and yet they are still smaller than both the reanalysis dataset and the model, this only serves to further highlight the difference between ORAS5, GFDL-OM4, and reality.

\subsubsection{Eddy Shape} \label{sec:shape}
\begin{figure}[htbp]
        \centering
        \includegraphics[width=\linewidth]{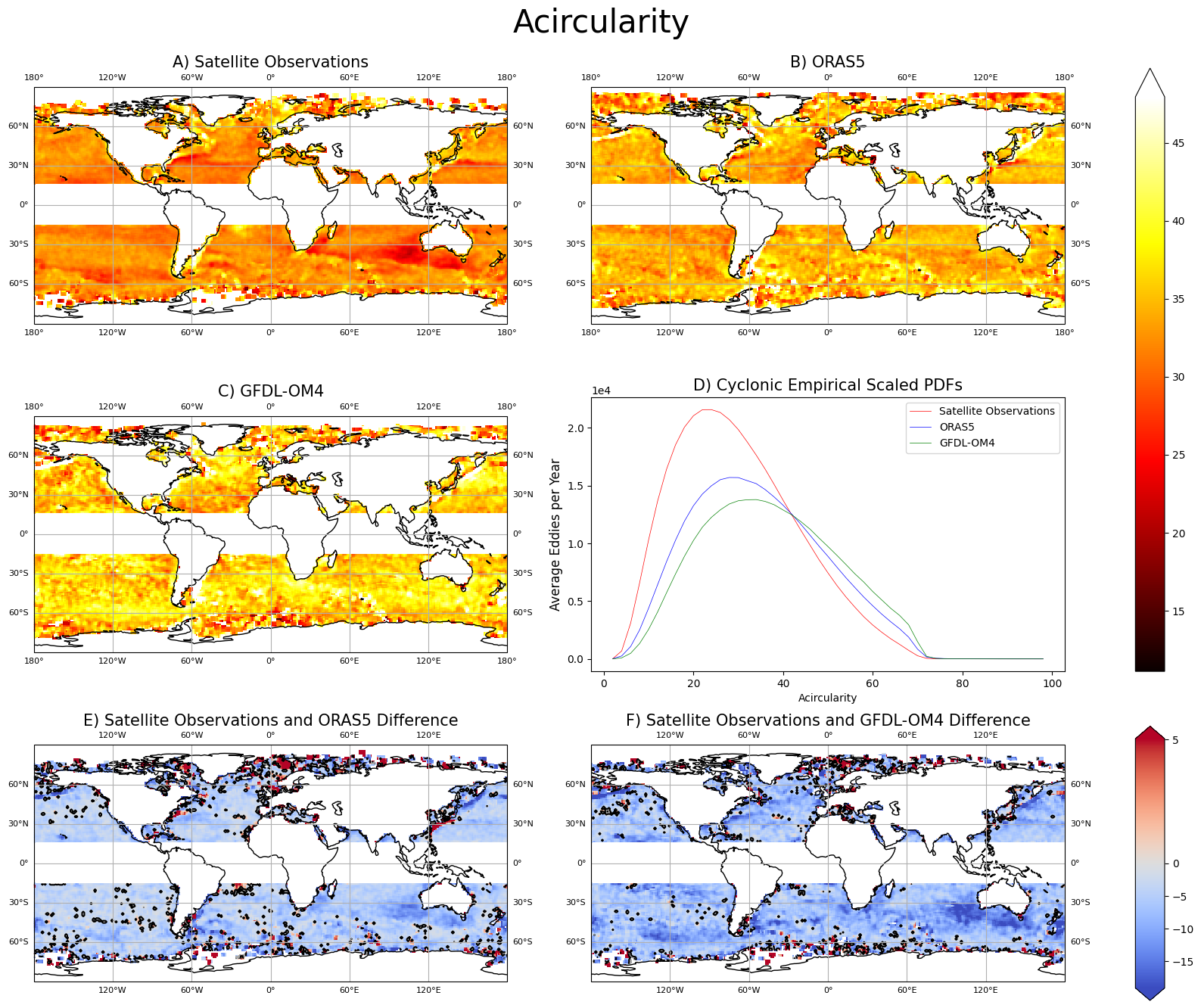}
        \caption{Cyclonic Eddy Acircularity.  (A-C) display the local mean maps of acircularity for ADT, ORAS5, and GFDL-OM4. (D) shows the scaled empirical global PDFs of acircularity for ADT, ORAS5, and GFDL-OM4. (E-F) show the differences between ADT and ORAS5 maps and the differences between ADT and GFDL-OM4 maps for acircularity. Areas that are grayed out do not have differences that are statistically significant. \label{fig:shape}}
    \end{figure}

    The acircularity of an eddy is a measure of how much the shape of an eddy deviates from a circle. 
    Acircularity is calculated as 100 times the ratio between the areal sum deviations of the contour from its best fit circle and the area of this best fit circle \citep{meta3.1}.
    This is the value of the calculation performed by the shape error test during the detection step of PET. 
    An acircularity of zero implies that the eddy shape is a circle.
    An acircularity of 30-40 generally indicates a more oval or ellipse-shaped eddy. 
    A larger acircularity (50+) indicates a more irregular eddy shape.
    The global scaled PDFs in Fig.\ \ref{fig:shape}d show that ORAS5 and GFDL-OM4 are missing more-circular eddies (i.e.\ eddies with low acircularity) and have an excess of irregularly-shaped eddies with an acircularity greater than $\approx 45$.
    
    We next consider the global distribution of acircularity for each population of eddies, independently from the total number of eddies. 
    For cyclonic eddies in ADT, ORAS5, and GFDL-OM4, the global mean acircularities are $29.57$, $34.19$, and $36.97$, the medians are $28.0$, $33.0$, and $36.0$, and the global standard deviations for acircularities are $13.37$, $14.19$, and $14.47$, respectively.
    Similar to eddy area, the average acircularity and standard deviation for both GFDL-OM4 and ORAS5 are larger than observations.
    This is consistent with the shapes of the PDFs shown in Fig.\ \ref{fig:shape}, where the global PDFs of acircularity from ORAS5 and GFDL-OM4 are more skewed towards large values.
    Overall, eddies in GFDL-OM4 and ORAS5 are less circular than eddies in ADT.
    
    Globally, we know that the distributions of acircularity in ORAS5 and GFDL-OM4 have more spread and a propensity towards less circular eddies. 
    Spatially, similar problems are present.
    Both ORAS5 and GFDL-OM4 have a larger acircularity than ADT almost everywhere. 
    There are not many regions where ORAS5 or GFDL-OM4 have correctly-shaped eddies.
    ORAS5 at a glance does exhibit some similar broader trends to ADT with having a higher acircularity near coastlines and lower acircularity in the open ocean. 
    However, ORAS5 fails to accurately represent more fine-scale local behavior.
    Notable areas where this happens are near the Australian Bight and in most of major currents.
    GFDL-OM4 does a worse job at capturing the correct behavior than ORAS5. 
    Like with the previous quantities these differences are statistically significant almost everywhere.
    
    This inability to accurately represent the spatial distribution of the mean acircularity is confirmed by the anomaly cosine correlation and RMSE for these maps. 
    The anomaly cosine correlation and RMSE between ADT and ORAS5 is $0.33$ and $4.83$ respectively. 
    The anomaly cosine correlation and RMSE between ADT and GFDL-OM4 is $0.22$ and $6.45$.
    The spatial distribution of the mean acircularity in both ORAS5 and GFDL-OM4 is substantially different than observed eddies in the ADT field.
    While ORAS5 outperforms GFDL-OM4 in this regard, both datasets are not accurate at capturing eddy shape in many areas of the globe.

\subsubsection{Eddy Speed Average} \label{sec:speed}
\begin{figure}[htbp]
    \centering
    \includegraphics[width=\linewidth]{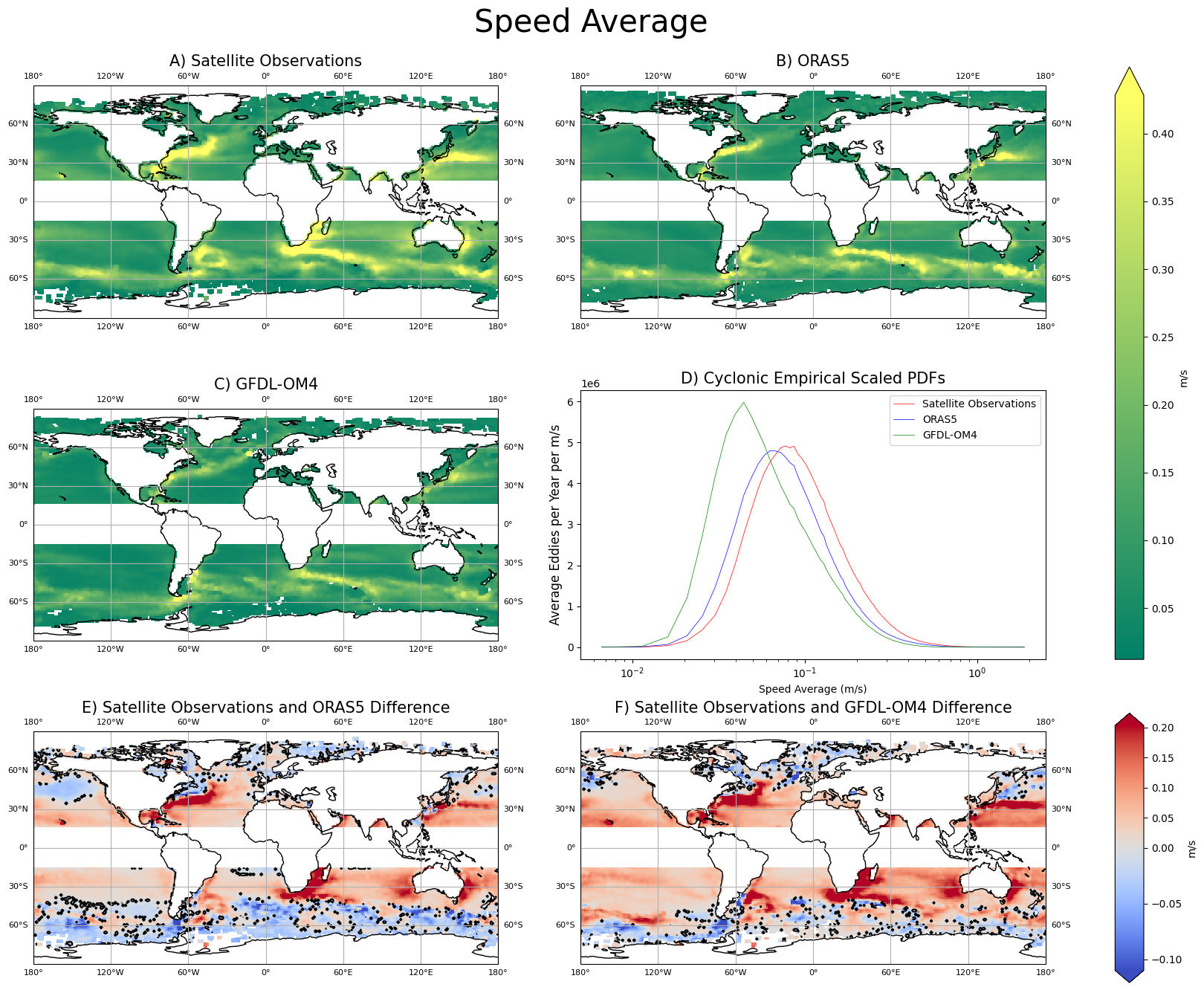}
    \caption{Cyclonic Eddy speed average.  (A-C) display the spatial maps of mean speed average for ADT, ORAS5, and GFDL-OM4. (D) shows the scaled empirical global PDFs of speed average for ADT, ORAS5, and GFDL-OM4. (E-F) show the differences between ADT and ORAS5 maps and the differences between ADT and GFDL-OM4 maps for speed average. Areas that are grayed out do not have differences that are statistically significant.\label{fig:speed}}
\end{figure}

    Eddy speed average is the average geostrophic speed along the contour defining the speed radius. Eddy speed average is related to the rotational speed of an eddy.
    The PDFs in Fig.\ \ref{fig:speed}d show that despite having substantially fewer eddies overall, GFDL-OM4 has an excess of slowly-rotating eddies with rotational speeds less than about 5 cm/s.
    In contrast, ORAS5 has a much better distribution of eddy rotational speeds. It has a little too many eddies with speeds below 5 cm/s and a little too few eddies with speeds above 5 cm/s.
    While the scaled PDF of ORAS5 could be corrected by shifting the entire distribution towards higher speeds, correcting GFDL-OM4 would require both removing lower speed eddies and adding new eddies with rotational speeds.  

    We next consider the global distribution of eddy rotational speeds for each population of eddies, independently from the total number of eddies. 
    The global mean speed averages for ADT, ORAS5, and GFDL-OM4 are $0.16$ m/s, $0.13$ m/s, and $0.11$ m/s, the medians are $0.12$ m/s, $0.10$ m/s, and $0.08$ m/s, and the standard deviations are $0.12$ m/s, $0.10$ m/s and $0.08$ m/s, respectively.
    Both ORAS5 and GFDL-OM4 have a lower speed average than ADT globally, consistent with the global PDF.

    In the spatial maps in Fig.\ \ref{fig:speed}a-c, the spatial structure of mean rotation speed in ORAS5 and GFDL-OM4 both broadly follow ADT,  with higher-speed eddies in the major current regions (ACC, Kuroshio extension, Gulf Stream, etc.).
    This is supported by the relatively-high anomaly cosine correlations of $0.78$ for ORAS5 and $0.68$ for GFDL-OM4.
    However, many of the smaller scale features of the spatial structure are wrong in both ORAS5 and GFDL-OM4.
    Eddies in ORAS5 and GFDL-OM4 tend to be too slow in the major current regions.
    Similar to amplitude, eddies in GFDL-OM4 tend to be too fast in the eastern half of the NAC.
    Lastly, eddies tend to also be too slow outside of these regions, most notably in the open ocean and particularly in the southern Indian ocean. 
    These differences are reflected by the RMSE values of $0.05$ m/s for ORAS5 and $0.07$ m/s for GFDL-OM4.
    As shown in Fig.\ \ref{fig:speed}e-f, outside of some isolated regions at high latitudes, these differences are statistically significant almost everywhere.

    ORAS5 not only is closer globally to the observed speed average distribution, but also better represents the local distribution of speed average when compared to GFDL-OM4. 
    However, ORAS5 still has room for improvement and seems to be missing many of the medium-speed eddies that are present in much of the open ocean. 

\subsection{Trajectory-Level Statistics}
In this section we transition to looking at statistics of eddy trajectories, recalling that we only consider trajectories which are composed of a minimum of 42 DED's, i.e.\ trajectories that are at least 6 weeks long.
In the following sections we will discuss eddy birth, azimuth, lifetime, displacement, and distance traveled.
\subsubsection{Eddy Birth} \label{sec:birth}
\begin{figure}[ht]
    \centering
    \includegraphics[width=\linewidth]{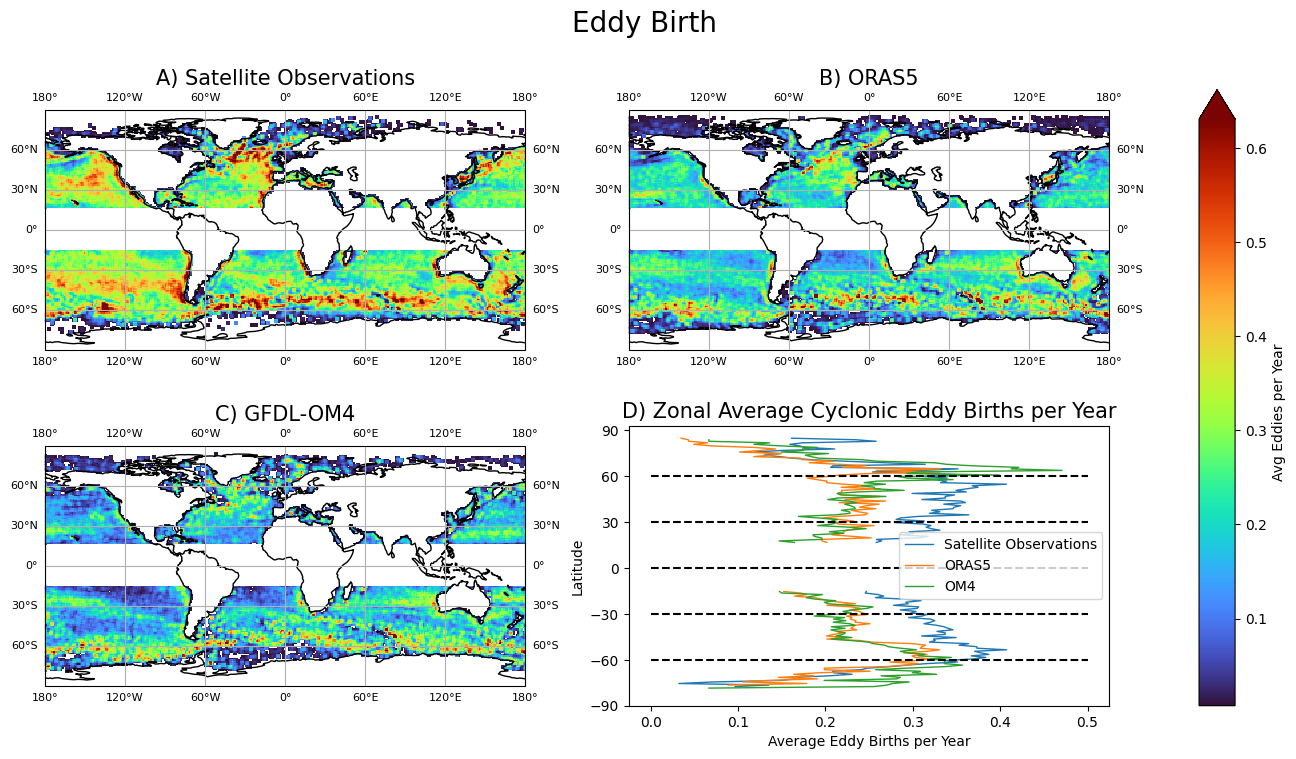}
    \caption{Cyclonic Eddy Births. (D) is the zonal average cyclonic eddy births per year. (A-C) depict the eddy births per year within each overlapping $3^\circ \times 3^\circ$ longitude-latitude box for ADT, ORAS5, and GFDL-OM4. \label{fig:birth}}
\end{figure}

We define the birth of an eddy to be the location of its first DED.
The average number of eddy births per year is displayed in Fig.\ \ref{fig:birth}. 
Calculating the average number of eddy births per year per $1^\circ\times 1^\circ$ box directly results in a noisy field. 
We smooth the results by by finding the average number of eddy births per year within a $3^\circ \times 3^\circ$ longitude-latitude box and then dividing this count by 9 to get the average number of eddy births per year per $1^\circ\times 1^\circ$ box. We then do this for all overlapping $3^\circ \times 3^\circ$ boxes on a $1^\circ$ grid.

In line with the results from Section \ref{sec:glob_stats}, the number of eddy births is lower almost everywhere in both ORAS5 and GFDL-OM4 than in ADT. 
ORAS5 and GFDL-OM4 are missing a large number of eddy births in the ACC, the north Atlantic, and north Pacific. 
They are also missing a considerable number of eddies in the open ocean in the mid-latitudes. 
ORAS5 and GFDL-OM4 both exhibit a similar trend to ADT of having a small number of eddy births near the equator and having an increased number of eddy births up until roughly $60^\circ$ N and $60^\circ$ S.

As we did with census, we also plot the zonal average of eddy birth in Fig.\ \ref{fig:birth}d. 
Just taking into account the reduced number of eddy trajectories in ORAS5 and OM4, one might expect that ORAS5 and OM4 to have uniformly less eddy births across all latitudes.  
However, similar to the trend observed with eddy census, when compared to ADT, ORAS5 and GFDL-OM4 have a similar number eddy births at high latitudes and fewer eddy births between $60^\circ$ S and $60^\circ$ N. 
While ORAS5 and GFDL-OM4 do have a similar number of eddy births at high latitudes when compared to ADT, compared to the total number of eddies generated by ORAS5 and GFDL-OM4, a disproportionally large number of them are generated at high latitudes.

When comparing to ADT, the anomaly cosine correlation of ORAS5 is $0.58$ and RMSE is $0.17$ eddy births per year.
The anomaly cosine correlation of GFDL-OM4 is $0.47$ and RMSE is $0.19$ eddy births per year. 
As was the case with many of the previous sections, ORAS5 does outperform GFDL-OM4 in both anomaly cosine correlation and RMSE. 
However, both ORAS5 and GFDL-OM4 fail to accurately capture much of the regional behavior of eddy birth.

\subsubsection{Eddy Azimuth} \label{sec:angle}
\begin{figure}[ht]
        \centering
        \includegraphics[width=\linewidth]{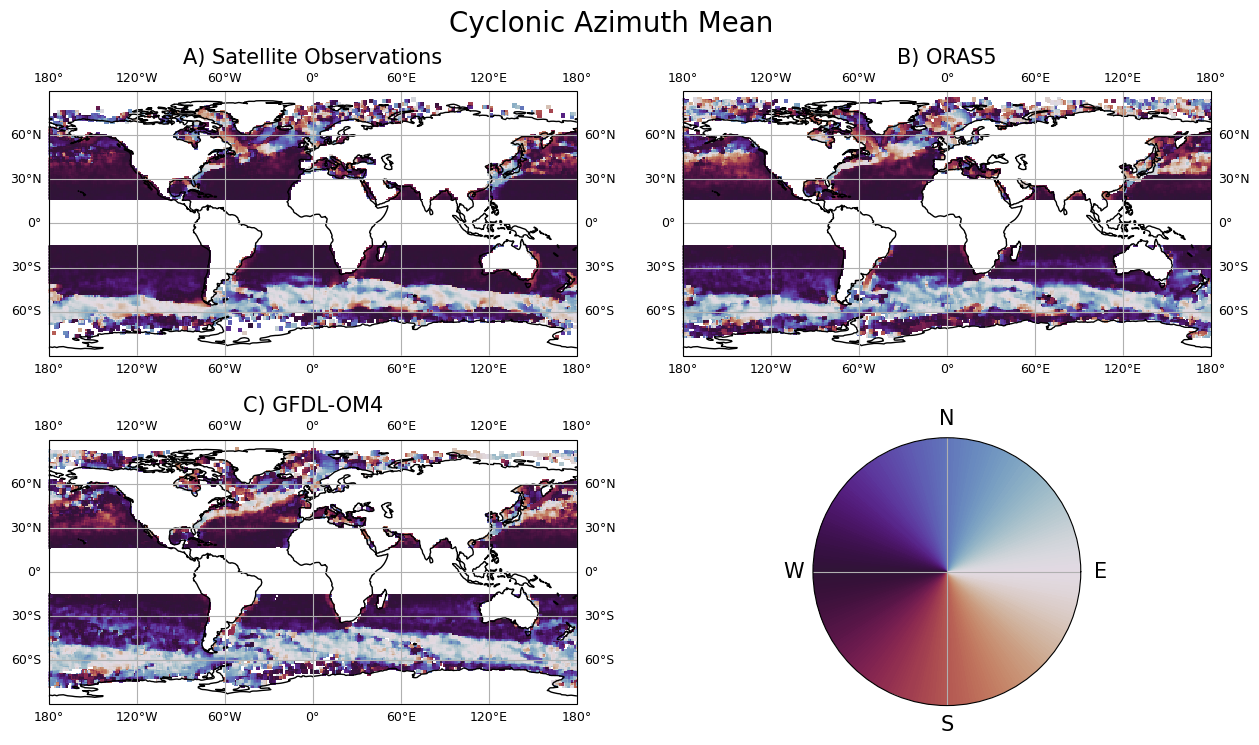}
        \caption{Cyclonic Eddy Azimuth Mean. (A-C) display the circular mean for eddy cyclonic eddy azimuth for ADT, ORAS5, and GFDL-OM4.\label{fig:angle_mean_cyc}}
    \end{figure}

\begin{figure}[ht]
        \centering
        \includegraphics[width=\linewidth]{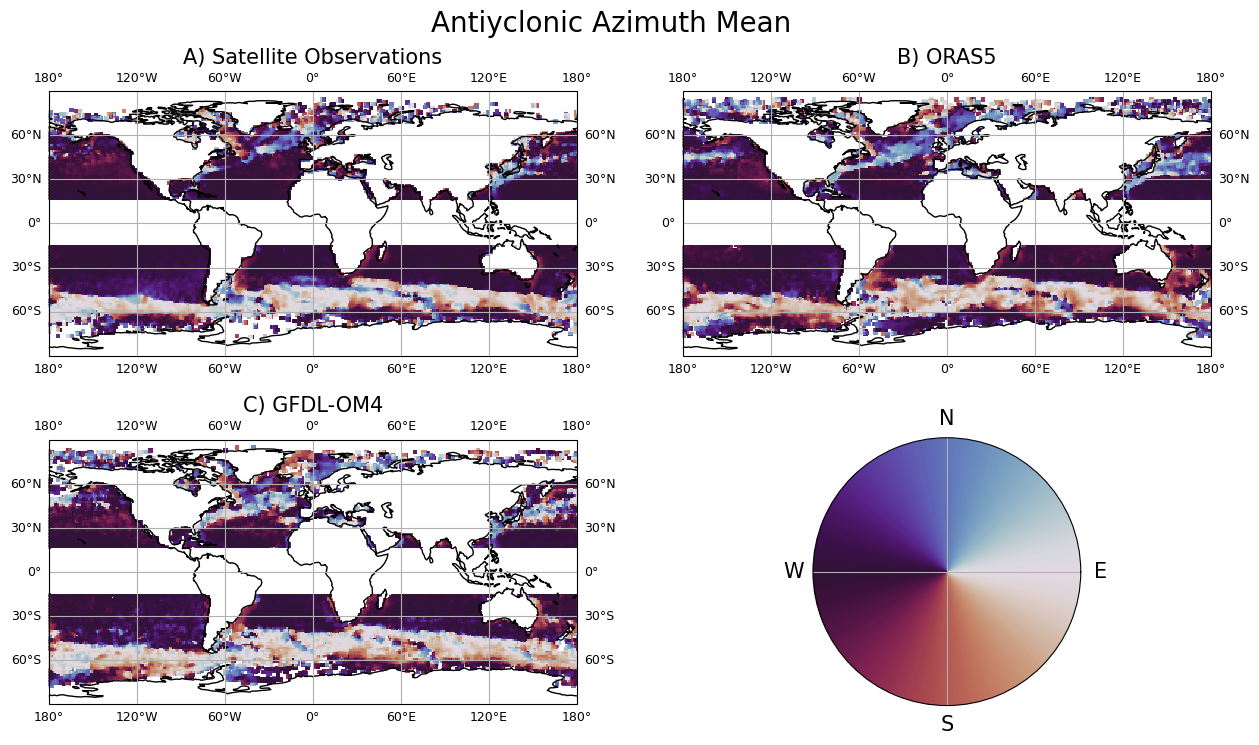}
        \caption{Anticyclonic Eddy Azimuth Mean. (A-C) display the circular mean for eddy anticyclonic eddy azimuth for ADT, ORAS5, and GFDL-OM4. \label{fig:angle_mean_ac}}
    \end{figure}

    We define Eddy Azimuth as the angle of a great-circle section connecting the start and end of an eddy trajectory, where the angle is measured in radians counterclockwise with respect to due east. 
    For example, if an eddy propagated directly eastward it would have an angle of $0$ radians.
    If an eddy propagated directly westward it would have an angle of $\pi$ radians. 
    The circular mean for angular data is defined as 
   \begin{equation}
    \text{Arg}\left(\frac{1}{n}\sum_{k=1}^ne^{ix_k}\right)
   \end{equation}
    where $i$ is the imaginary unit, $x_k$ are the given angle observations in radians, and $\text{Arg}(z)$ gives the principal value of the argument of $z$ restricted to $[0,2\pi]$. 
    Figs.\ \ref{fig:angle_mean_cyc} and \ref{fig:angle_mean_ac} show the circular mean for all cyclonic and anticyclonic eddies born in each overlapping $3^\circ \times 3^\circ$ box. 
    
    Consistent with \citep{chelton}, globally almost all eddies on average propagate westward. 
    There are a few notable exceptions to the trend which occur in the North Atlantic subpolar gyre and Norwegian Sea, and especially along the ACC.
    These broad trends are shared by ADT, ORAS5, and GFDL-OM4.
    Cyclonic eddies in both ORAS5 and GFDL-OM4 have a primarily eastward propagation in the Gulf Stream, NAC, and the Kuroshio extension, which is not the case for the observed ADT eddies. 
    Between $30^\circ $ N and $60^\circ$ N in the Pacific Ocean there is less westward propagation and more south-west propagation of cyclonic eddies in ORAS5 and GFDL-OM4 when compared to ADT.
    Similarly between $30^\circ $ S and $60^\circ$ S there tends to be more northward propagation of cyclonic eddies in ORAS5 and GFDL-OM4 than in ADT.
    Cyclonic eddies in the ACC in ORAS5 and GFDL-OM4 tend to primarily propagate east or northeast which differs from observed eddies in ADT which frequently propagate southeast as well. 
    We do not present results on the significance of the differences for eddy azimuth. 
    We investigated using the Two-sample Kuiper Test to test whether or not the angular data comes from the same distribution. 
    However, computing the p-values for small samples is quite approximate.
    As such, we have elected not to present these results due to concerns over the accuracy of these p-values.
    
    Similar trends are observed in the anticyclonic azimuth means in Fig.\ \ref{fig:angle_mean_ac}. In ORAS5 and GFDL-OM4 anticyclonic eddies in the Gulf Stream and Kuroshio extension propagate north/northeast as opposed to east like their cyclonic counterparts. Also the anticyclonic eddies in ORAS5 and GFDL-OM4 tend to propagate more southeast in the ACC than their counterparts in ADT.
    
\begin{figure}[ht]
        \centering
        \includegraphics[width=\linewidth]{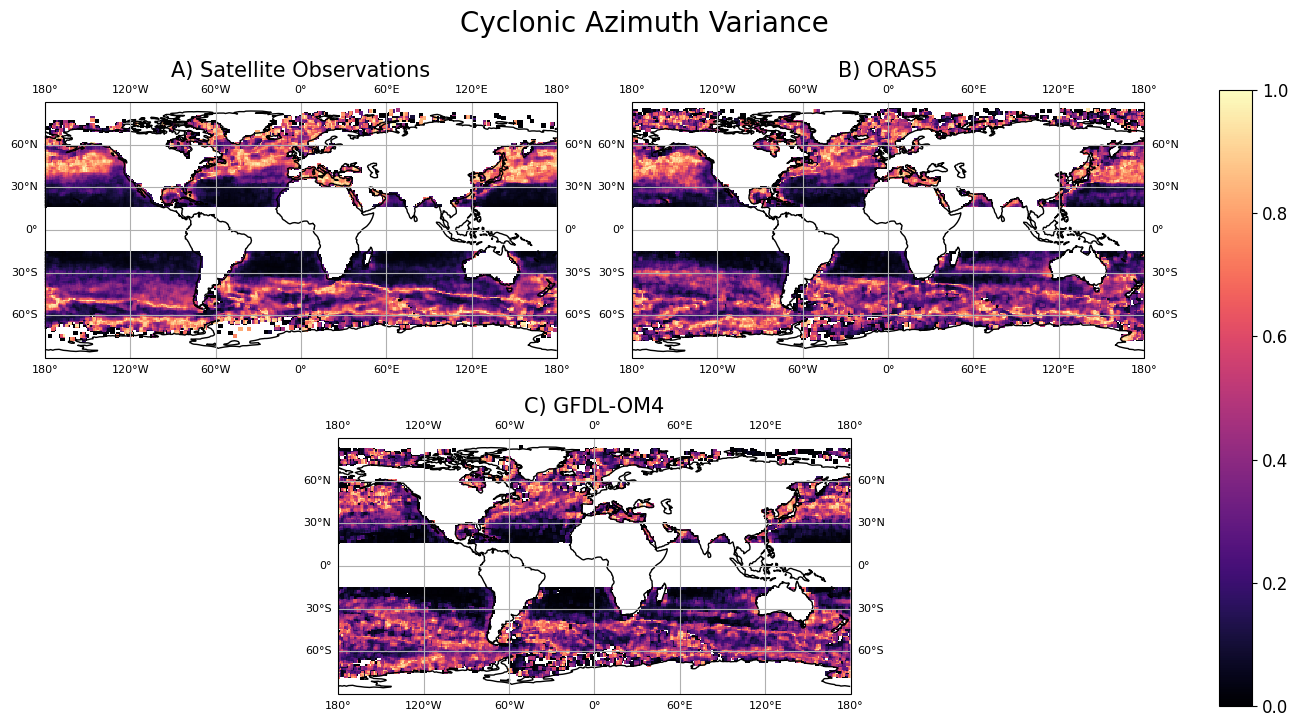}
        \caption{Cyclonic Eddy Azimuth Variance.  (A-C) display the circular variance for eddy anticyclonic eddy azimuth for ADT, ORAS5, and GFDL-OM4. \label{fig:angle_var}}
    \end{figure}

    Circular variance is defined as 
    \begin{equation}
    1 - \left|\frac{1}{n}\sum_{k=1}^ne^{ix_k}\right|.
    \end{equation}
    where $x_k$ are the given angle observations in radians. Circular variance is between 0 and 1; a uniform distribution of angles has circular variance 1. The circular variance for all overlapping $3^\circ\times 3^\circ$ boxes is displayed for cyclonic eddies in Fig.\ \ref{fig:angle_var}.
    The circular variance for ADT is low in the mid-latitudes where eddies propagate primarily west and  then increases towards 1 above $30^\circ$ N and below $30^\circ$ S. 
    GFDL-OM4 and ORAS5 exhibit similar broader trends however they don't transition as quickly from areas of low circular variance to areas of high circular variance.
    Notably, both model-based datasets have higher circular variance in the southern Indian ocean. 
    Lastly, both ORAS5 and GFDL-OM4 exhibit different patterns of high variance in both the ACC and at higher latitudes, which could be associated with biases in the position and strength of the eastward mean currents in these regions.
    
    To quantify the similarity in these maps for eddy azimuth we use the correlation coefficient for circular data proposed by \citep{angle_corr}.
    Given a random sample $p_i = (\theta_i,\phi_i), i = 1,...,n$ from a distribution on the torus $(\Theta, \Phi)$, the estimator for the correlation coefficient for the circular data is defined by
    \begin{equation} \label{eq:angle_sim}
        \hat{\rho}_T = \frac{\sum \sin(\theta_i  - \theta_j) \sin(\phi_i - \phi_j)}{\sqrt{\sum (\sin(\theta_i  - \theta_j))^2} \sqrt{\sum (\sin(\phi_i  - \phi_j))^2}}
    \end{equation}
    where $\sum = \sum_{i=1}^{n-1} \sum_{j=i+1}^n$.
    This circular correlation coeffiecient is between -1 and 1 and has a similar interpretation to the Pearson correlation coefficient. 
    If the circular correlation is 1, then $\Theta \equiv \Phi + \alpha_0 \pmod{2\pi}$ where $\alpha_0$ is a fixed arbitrary direction. 
    If the circular correlation is -1, then $\Theta \equiv -\Phi + \alpha_0 \pmod{2\pi}$.
    We define error for our circular data to be 
    \begin{equation} \label{eq:anlge_rmse}
        \frac{1}{n}\sum_{i=1}^n \sqrt{2-2\cos(\theta_i - \phi_i)}.
    \end{equation}
    which is derived from projecting $\theta_i$ and $\phi_i$ onto the unit circle and calculating the Euclidean distance between both points. We then average this across all points in the dataset. Both the circular correlation and error for the mean and the anomaly cosine correlation and RMSE for the variance are displayed in Table \ref{tab:angle_sim}.
\begin{table}[ht]
\centering
\footnotesize
\begin{tabular}{lcccccccc}
\hline
Metric& \multicolumn{2}{c}{Cyc: ADT vs. ORAS5} & \multicolumn{2}{c}{Cyc: ADT vs. GFDL-OM4} & \multicolumn{2}{c}{Ac: ADT vs. ORAS5} & \multicolumn{2}{c}{Ac: ADT vs. GFDL-OM4} \\
 & Sim & Error & Sim & Error & Sim & Error & Sim & Error \\
\hline
Circular Mean & 0.32 & 0.55 & 0.28 & 0.61 & 0.30 & 0.54 & 0.28 & 0.56 \\
Circular Variance  & 0.65 & 0.15 & 0.57 & 0.16 & 0.55 & 0.16 & 0.49 & 0.17 \\
\hline
\end{tabular}
\caption{Comparison of Angle for ADT against ORAS5 and GFDL-OM4. For circular mean, sim refers to the correlation coefficient for circular data defined in \ref{eq:angle_sim} and  error refers to the error defined in \ref{eq:anlge_rmse}. For circular variance, sim refers the anomaly cosine correlation and error refers to RMSE.\label{tab:angle_sim}}
\label{tab:adt_comparison}
\end{table}

The circular correlation for the the cyclonic circular mean is low, being $0.3181$ between ADT and ORAS5 and $0.286$ between ADT and GFDL-OM4. 
While the broad trends are similar between all three datasets, ORAS5 and GFDL-OM4 struggle at accurately capturing the spatial distribution of the circular mean.
Both ORAS5 and GFDL-OM4 have a similar error when compared to ADT, with the error being $0.5516$ and $0.6050$ respectively.
This chordal distance on the unit circle corresponds to a central angle of 0.61 radians, or about 35 degrees.
Unlike the previous eddy characteristics that we have presented, for eddy azimuth ORAS5 does not have as large of an advantage over GFDL-OM4. 
This could suggest that assimilating SSH data at a $1/4^\circ$ resolution does not provide a significant benefit to accurately simulating eddy azimuth. 

The anomaly cosine correlation for cyclonic circular variance is $0.65$ between ADT and ORAS5 and $0.57$ between ADT and GFDL-OM4.
The RMSE between ADT and ORAS5 is $0.1497$ and is $0.1663$ between ADT and GFDL-OM4.
Similar to circular mean, there is little difference between ORAS5 and GFDL-OM4 when comparing their spatial patterns of circular variance to ADT, as measured by anomaly correlation and RMSE. 
The combination of moderate correlation coefficient and low errors suggests that while on average the magnitude of the differences in the variance in any given box is low, the actual spatial relationship of circular variance is still not being fully captured. 

\subsubsection{Eddy Lifetime} \label{sec:life}
\begin{figure}[htbp]
        \centering
        \includegraphics[width=\linewidth]{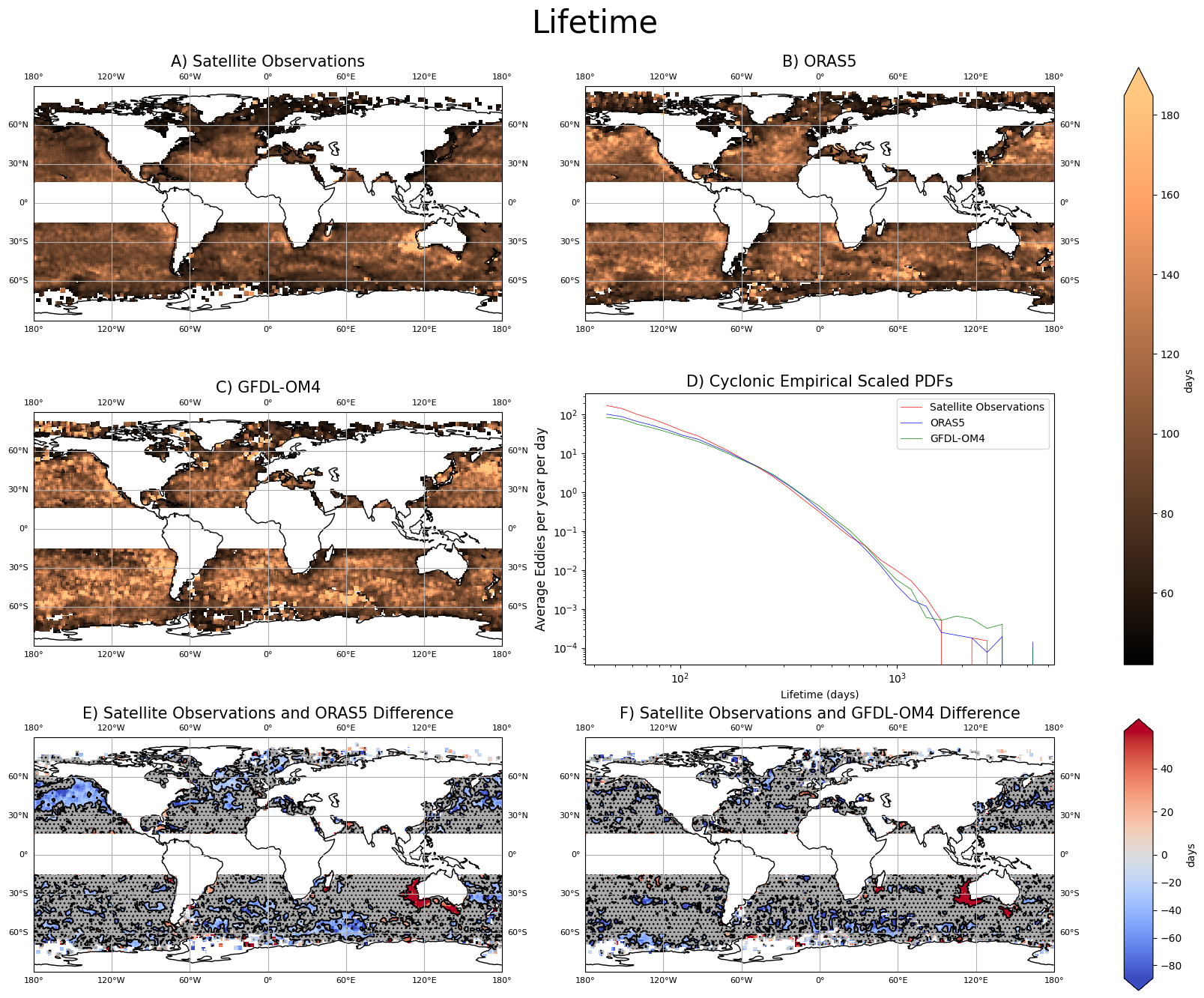}
        \caption{Cyclonic Eddy Lifetime.  (A-C) display the local mean maps of lifetime for ADT, ORAS5, and GFDL-OM4. (D) shows the scaled empirical global PDFs of lifetime for ADT, ORAS5, and GFDL-OM4. (E-F) show the differences between ADT and ORAS5 maps and the differences between ADT and GFDL-OM4 maps for lifetime. Areas that are grayed out do not have differences that are statistically significant.\label{fig:life}}
    \end{figure}

    Eddy lifetime is defined to be how long an eddy lives in days.    
    The global scaled PDFs displayed in Fig.\ \ref{fig:life}d show both ORAS5 and GFDL-OM4 have too few eddies with lifetimes less than  $\approx 100$ days.
    This lack of short-lived eddies in the models might be related to a lack of stochastic processes generating eddies.
    Coupled interactions between ocean mesoscale eddies and the atmosphere are known to damp eddies \citep{duhaut2006wind,xu2016work,yang2019mesoscale,renault2023modulation}; the lack of short-lived eddies in these models could perhaps also be related to an imperfect representation of such coupled processes in these forced (i.e.\ uncoupled) ocean models.
    All three datasets have a similar number of eddies that live for approximately several hundred days. 
    However, both ORAS5 and GFDL-OM4 have more eddies that live thousands of days than ADT.
    It should also be noted here that unlike the ADT and ORAS5 datasets that span 31 years, the GFDL-OM4 dataset is only 15 years long. 
    Despite this shorter time range it still has the longest lived eddies.
    We expect that if longer GFDL-OM4 runs were available, they would have even more long-lived eddies.

    We next consider the global distribution of eddy lifetimes for each population of trajectories, independently from the total number of trajectories.
    The global mean cyclonic lifetimes for ADT, ORAS5, and GFDL-OM4 are $93.5$ days, $104.7$ days, and $109.8$ days, the medians are $71$ days, $79$ days, and $82$ days, and the standard deviations are $69.2$ days, $80.03$ days, and $90.21$ days, respectively.
    Eddies in ORAS5 and GFDL-OM4 live longer and have substantially more variability in lifetime when compared to observed eddies in ADT. 
    
    The spatial maps for mean cyclonic eddy lifetime are displayed in Fig.\ \ref{fig:life}. Although differences in the mean lifetime are evident throughout, many of these differences are not statistically significant. For example, eddies born off the west coast of South America in ORAS5 and GFDL-OM4 live longer than their counterparts in ADT, but there are not enough of these eddies for the difference in mean lifetime to be statistically significant.
    When testing the significance of these results in  Fig. \ \ref{fig:life}e-f, we fail to reject the null hypothesis that the means are equal in most areas.
    It should be noted that for all of the trajectory level statistics the sample size is much lower in each $3^\circ\times 3^\circ$ longitude-latitude box.
    This small sample size combined with the high variance of these quantities reduces the power of the tests we perform.
    Considering all locations regardless of statistical significance, we find that the anomaly cosine correlation of ORAS5 to ADT is $0.23$ with an RMSE of $28.5$ days, while the anomaly cosine correlation of GFDL-OM4 to ADT is $0.09$ with an RMSE of $54.6$ days.
    Data from ORAS5 is available over a longer time period than GFDL-OM4, so there are more regions with statistically-significant differences in ORAS5, despite the fact that the differences are generally larger in GFDL-OM4.
    ORAS5 has longer-lived eddies than ADT in the subtropical north Pacific, the North Atlantic Current and parts of the subpolar North Atlantic, and in parts of the Southern Ocean.
    Both ORAS5 and GFDL-OM4 have eddies with too-short lives off the west and south coasts of Australia.

\subsubsection{Eddy Displacement} \label{sec:disp}
\begin{figure}[htbp]
        \centering
        \includegraphics[width=\linewidth]{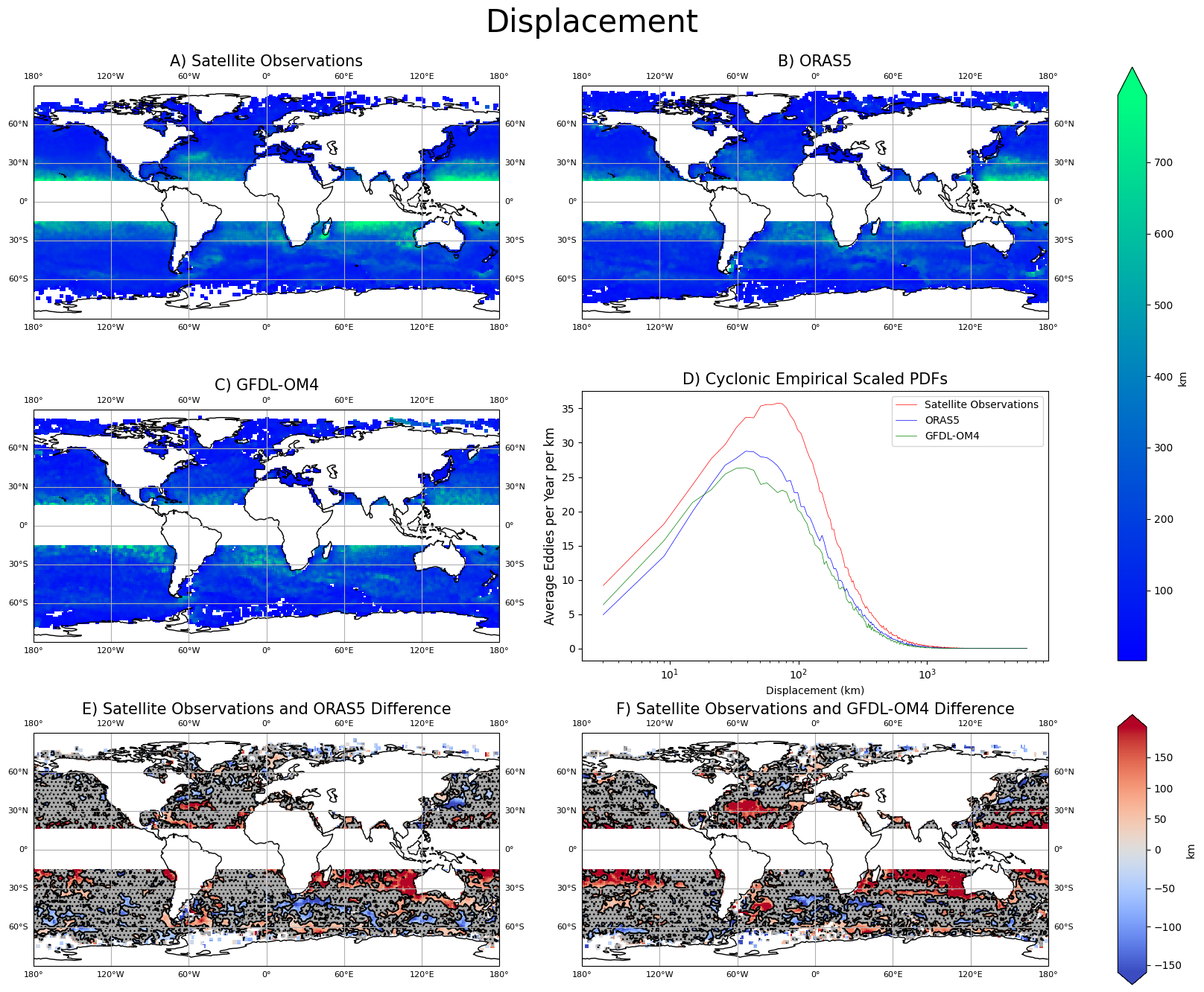}
        \caption{Cyclonic Eddy Displacement. (A-C) display the spatial maps of mean displacement for ADT, ORAS5, and GFDL-OM4. (D) shows the scaled empirical global PDFs of displacement for ADT, ORAS5, and GFDL-OM4. (E-F) show the differences between ADT and ORAS5 maps and the differences between ADT and GFDL-OM4 maps for displacement. Areas that are grayed out do not have differences that are statistically significant.\label{fig:disp}}
    \end{figure}

    We calculate eddy displacement to be the geodesic distance in kilometers between an eddy's birth location and death location.
    The global scaled PDFs in Fig.\ \ref{fig:disp}d show that ORAS5 and GFDL-OM4 are missing eddies across the entire range of displacements, primarily near the peak displacement of about 70 km.

    We next consider the global distribution of eddy displacements for each population of trajectories, independently from the total number of trajectories.
    The global mean displacements for cyclonic eddies in ADT, ORAS5, and GFDL-OM4 are $206.4$ km, $190.1$ km, and $175.1$ km, the medians are $132.8$ km, $126.2$ km, and $119.4$ km, and the standard deviations are $242.5$ km, $206.6$ km, and $182.9$ km, respectively.
    It is interesting  that the typical displacement for all eddies living at least 6 weeks or longer is only 70 -- 200 km, depending on the data set and on the statistic of interest, which is the same order of magnitude as a typical eddy diameter.
    Although the distribution is skewed towards larger displacements, many eddies only move a relatively short distance over their lifetimes.
    Compared to ADT, on average eddies in ORAS5 and GFDL-OM4 do not propagate far enough before they die, and there is less variability in the distribution of displacement.
    
    
    The spatial maps of mean eddy displacement in Fig.\ \ref{fig:disp}a-c all have a similar pattern. The main difference that ORAS5 and GFDL-OM4 share is a smaller displacement than ADT just outside the tropics. 
    ORAS5 and GFDL-OM4 are also both missing the higher-propagation eddies off of the west coast of Australia.
    These differences are only statistically significant in the subtropical South Pacific and Indian Oceans, particularly off the west coast of Australia.
    GFDL-OM4 also has a statistically significant difference in displacement in the middle of the North Atlantic and above and below the tropics in the Pacific.
    The anomaly cosine correlation for these maps of mean eddy displacement are $0.85$ and $0.75$ for ORAS5 and GFDL-OM4 respectively.
    The RMSE is $61.2$ km and $77.0$ km for ORAS5 and GFDL-OM4 respectively. 
    Both ORAS5 and GFDL-OM4 do a perform well at capturing the spatial patterns of eddy displacement, with ORAS5 performing slightly better and having a lower RMSE. 
    
\begin{figure}[htbp]
        \centering
        \includegraphics[width=\linewidth]{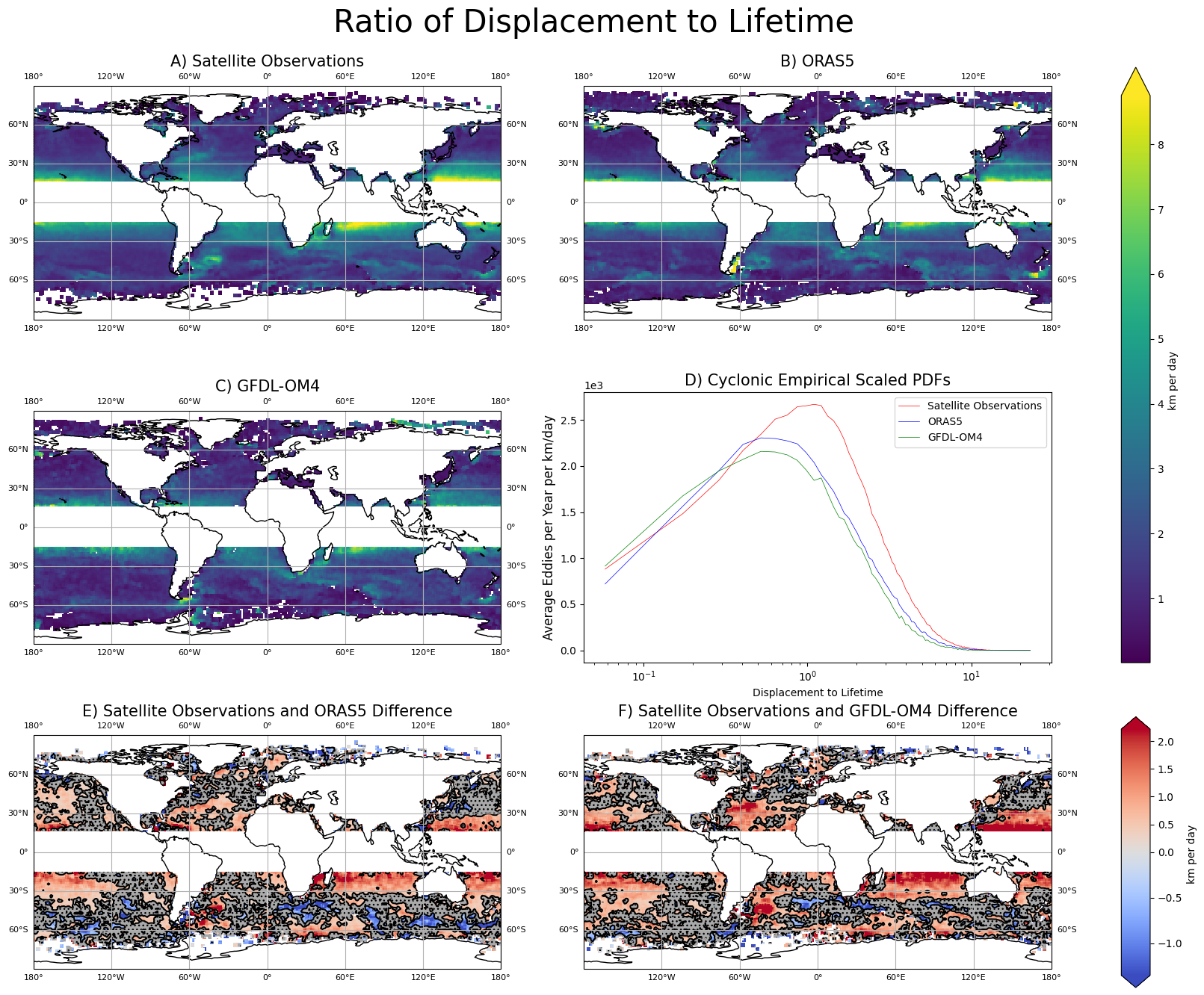}
        \caption{Cyclonic Ratio of Displacement to Lifetime. (A-C) display the spatial maps of the mean ratio of displacement to lifetime for ADT, ORAS5, and GFDL-OM4. (D) shows the scaled empirical global PDFs of the ratio of displacement to lifetime for ADT, ORAS5, and GFDL-OM4. (E-F) show the differences between ADT and ORAS5 maps and the differences between ADT and GFDL-OM4 maps for the ratio of displacement to lifetime. Areas that are grayed out do not have differences that are statistically significant.\label{fig:disp_life}}
    \end{figure}

    It should be noted that both eddy displacement and eddy distance (Section \ref{sec:dist}) depend heavily on eddy lifetime (Section \ref{sec:life}) as the longer an eddy lives, the further it is able to propagate. If eddy displacement were solely determined by lifetime, then we would expect eddies in ORAS5 and GFDL-OM4 to travel farther than in ADT, but the opposite is true, which suggests some deficiency in the representation of eddy dynamics in ORAS5 and GFDL-OM4. 
    One exception to this paradox is the area of eddy births off the west coast of Australia: the lifetimes and displacements of eddies born in this region in both ORAS5 and GFDL-OM4 are both too short. 
    To further examine the relationship between lifetime and displacement, we also present statistics on the ratio of displacement to lifetime, which we call `displacement velocity.' 
    The global scaled PDFs in Fig.\ \ref{fig:disp_life}d show that ORAS5 and GFDL-OM4 have slightly more eddies with a displacement velocity of approximately 30 km/day and are missing eddies with larger displacement velocities.

    Independent of the total number of trajectories, we find that the global means of displacement to lifetime for cyclonic eddies in ADT, ORAS5 and GFDL-OM4 are $2.24$ km/day, $1.98$ km/day, and $1.77$ km/day, the medians are $1.76$ km/day, $1.50$ km/day, and $1.37$ km/day, and the standard deviations are $1.80$ km/day, $1.72$ km/day, $1.49$ km/day, respectively. 

    The spatial mean maps in Fig.\ \ref{fig:disp_life}a-c are similar to those presented in Fig.\ \ref{fig:disp}. 
    There are large differences above and below the tropics, particularly in the Indian and Pacific ocean. 
    Fig.\ \ref{fig:disp_life}e-f shows that a substantial amount of the differences between ADT and ORAS5 and GFDL-OM4 are significant with most of the significant differences occurring between $30^\circ$ S and $30^\circ$ N.
    Considering the ratio of lifetime to displacement leads to more regions where the differences are statistically significant, as can be seen by comparing panels E-F) in figures \ref{fig:life}, \ref{fig:disp}, and \ref{fig:disp_life}.
    Throughout the subtropics, eddies in both ORAS5 and GFDL-OM4 have too low mean displacement velocities.

    Outside of these differences the overall spatial pattern of these maps is quite similar. 
    Compared to ADT, the anomaly cosine correlation of ORAS5 is $0.83$ and RMSE is $0.66$ km/day.
    The anomaly cosine correlation of GFDL-OM4 is $0.80$ and RMSE is $0.75$ km/day.
    While the pattern for displacement was already good, when accounting for lifetime, ORAS5 and GFDL-OM4 still are able to accurately capture the spatial distribution of displacement.

\subsubsection{Eddy Distance Traveled} \label{sec:dist}
\begin{figure}[htbp]
        \centering
        \includegraphics[width=\linewidth]{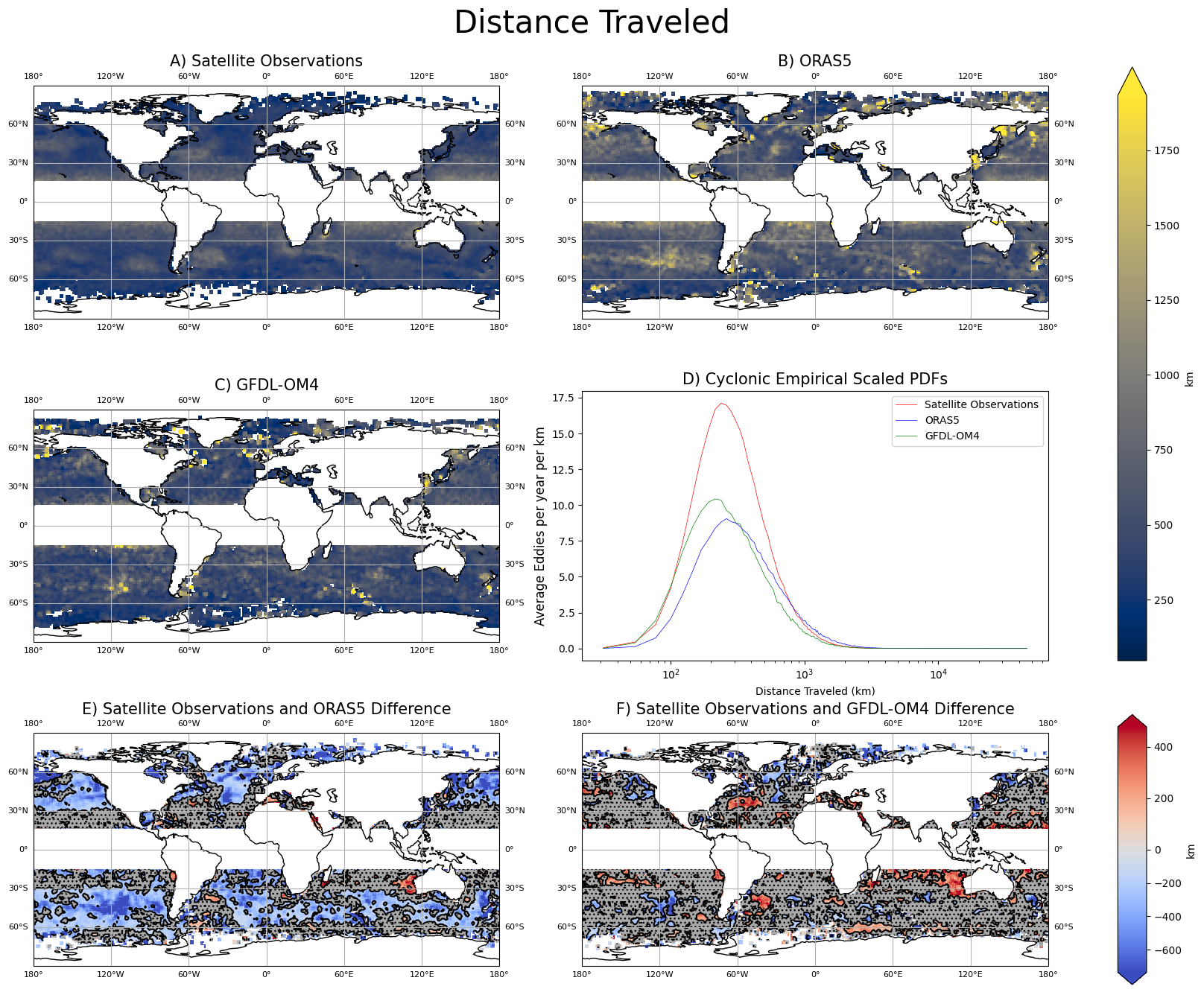}
        \caption{Cyclonic Eddy Distance Traveled. (A-C) display the spatial maps of mean distance traveled for ADT, ORAS5, and GFDL-OM4. (D) shows the scaled empirical global PDFs of distance traveled for ADT, ORAS5, and GFDL-OM4. (E-F) show the differences between ADT and ORAS5 maps and the differences between ADT and GFDL-OM4 maps for distance traveled. Areas that are grayed out do not have differences that are statistically significant.\label{fig:dist}}
    \end{figure}

    We define eddy distance traveled to be the total path length of an eddy trajectory in kilometers. 
    The path length of an eddy trajectory is calculated by summing the geodesic distance between each of an eddy's daily detections.
    The global scaled PDFs in Fig.\ \ref{fig:dist}d show that ORAS5 and GFDL-OM4 are primarily missing eddies whose distance traveled is close to the mode, around 120 km.
    GFDL-OM4 has about the right number of eddies with small distances traveled of about 100 km or less, while ORAS5 has too few of these eddies.
    While GFDL-OM4 has too few eddies with longer distances traveled, ORAS5 is more accurate, and has a slight excess of eddies with distances traveled on the order of 1,000 km or greater.

    Considering next population statistics independent of the total number of trajectories, we find that the global means of cyclonic distance traveled  for ADT, ORAS5, and GFDL-OM4 are $487.38$ km, $667.46$ km, and $509.02$ km, the medians are $379.31$ km, $488.89$ km, and $379.05$ km, and the standard deviations are $377.68$ km, $642.42$ km, and $512.70$ km, respectively.
    Unlike displacement, eddies in ORAS5 and GFDL-OM4 travel farther than they do in ADT and have more variability in the distance that they travel.
    Statistics involving distance traveled (including the ratio of displacement to distance traveled) are the only ones where GFDL-OM4 is closer to ADT than ORAS5.

    The spatial maps of mean distance traveled are displayed in Fig.\ \ref{fig:dist}. 
    In both ORAS5 and GFDL-OM4 eddies in the subtropics tropics travel farther than observed eddies in ADT, but these differences are mostly not statistically significant. 
    In both ORAS5 and GFDL-OM4 eddies travel farther at high latitudes, particularly in the ACC and near New Zealand.
    Both ORAS5 and GFDL-OM4 have hot spots where distances traveled are too large; the statistical significant plots in panels E-F) show that there are also `cold spots' where the distances traveled are too small.
    One exception is that eddies born off the west coast of Australia in ORAS5 typically don't travel far enough, which is consistent with the behavior for both lifetime and displacement shown above; the behavior of eddies in GFDL-OM4 in this region is similar.
    Although panel F) shows that we fail over a large region to reject the null hypothesis that the mean distances traveled in ADT and GFDL-OM4 are the same, a region larger than for ORAS5 in panel E), it should be remembered that there is only about half as much data for GFDL-OM4 as there is for ORAS5, so the power of the statistical significance test is weaker.
    Returning to the spatial structure shown in Fig.\ \ref{fig:dist}, we report that the anomaly cosine correlation of ORAS5 to ADT for distance traveled is $0.24$ with a RMSE of $283.3$ km. 
    The anomaly cosine correlation of GFDL-OM4 to ADT is $0.09$ with a RMSE of $335.2$ km. Neither ORAS5 nor GFDL-OM4 excels at capturing the local structure of distance traveled.

\begin{figure}[htbp]
        \centering
        \includegraphics[width=\linewidth]{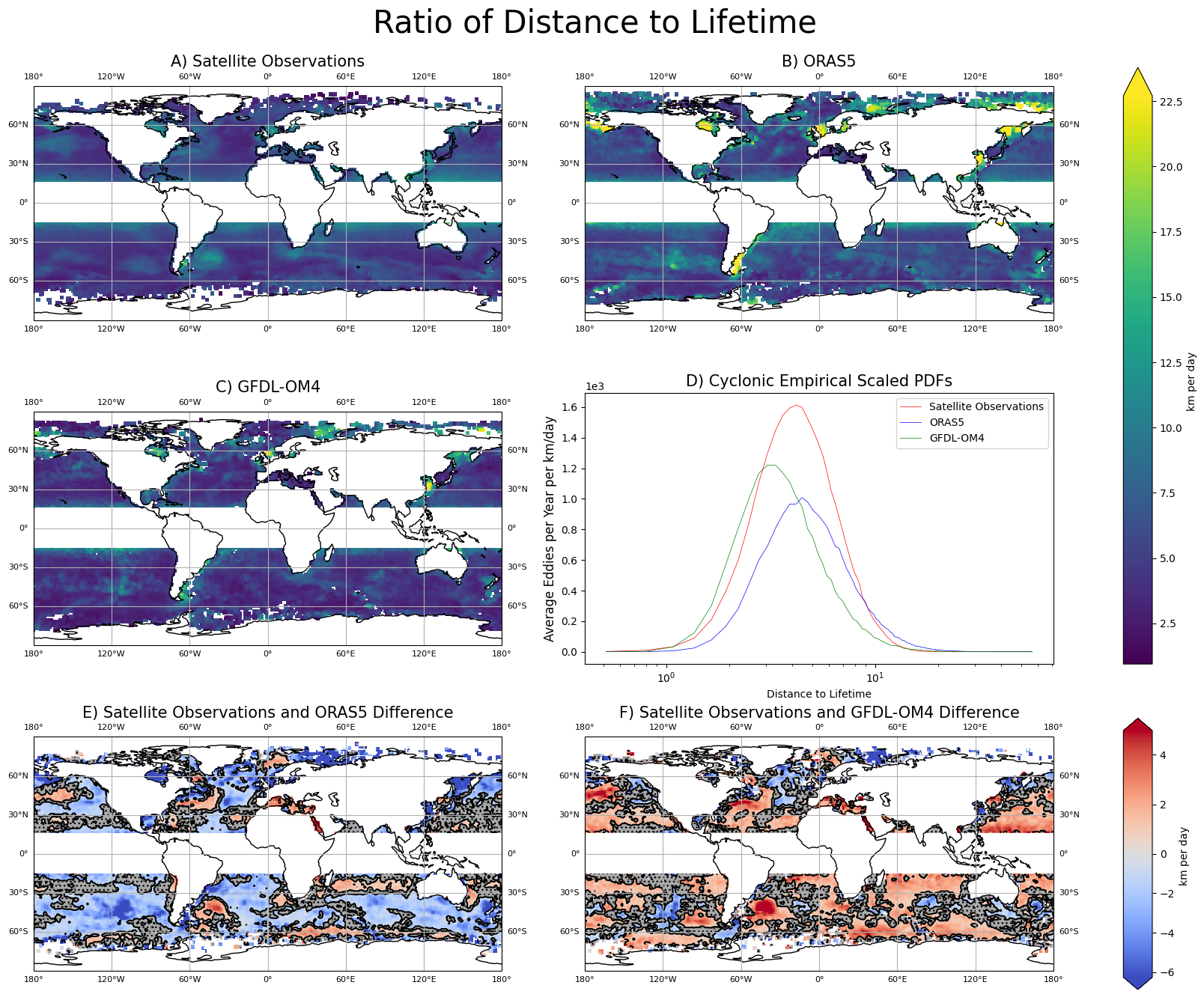}
        \caption{Ratio of Distance to Lifetime. (A-C) display the spatial maps of the mean ratio of distance to lifetime for ADT, ORAS5, and GFDL-OM4. (D) shows the scaled empirical global PDFs of the ratio of distance to lifetime for ADT, ORAS5, and GFDL-OM4. (E-F) show the differences between ADT and ORAS5 maps and the differences between ADT and GFDL-OM4 maps for the ratio of distance to lifetime. Areas that are grayed out do not have differences that are statistically significant.\label{fig:dist_life}}
    \end{figure}

    Distance, displacement, and lifetime are all related to each other, so to untangle the dependence somewhat we now present statistics on the ratio of distance to lifetime, which we call propagation speed.
    The global scaled PDFs of mean propagation speed in Fig.\ \ref{fig:dist_life}d show that the number of eddies with mean speeds of 1 to 2 km per day is approximately correct though slightly high in GFDL-OM4, which is primarily missing eddies with faster mean propagation speeds.
    In contrast, ORAS5 is missing eddies across the range of mean propagation speeds until one reaches speeds of about 10 km per day, at which point ORAS5 has more eddies.

    Considering next the distribution of mean propagation speed independently from the total number of trajectories, we find that the global cyclonic means of propagation speed for ADT, ORAS5, and GFDL-OM4 are 5.37 km/day, 6.50 km/day, and 4.87 km/day, the medians are 4.94 km/day, 5.63 km/day, and 4.15 km/day, and the standard deviations are 2.34 km/day, 3.63 km/day, and 2.82 km/day, respectively.
    As with displacement, considering the ratio of distance traveled to lifetime increases the area over which the differences in the means between ADT on one hand and ORAS5 and GFDL-OM4 on the other hand are statistically significant.
    Fig.\ \ref{fig:dist_life}e shows that the propagation speed of eddies in ORAS5 are generally too high, with some important exceptions including: the vicinity of the North and South Atlantic Currents, the subtropical Indian Ocean west of Australia, the Kuroshio Extension, and the Mediterranean and Red Seas, where the propagation speeds are too low.
    Fig.\ \ref{fig:dist_life}f shows that the propagation speed of eddies in GFDL-OM4 are quite different from ORAS5, being generally too low globally, with only a few exceptions, including the subpolar North Atlantic.
    Since the typical propagation speeds are a few kilometers per day, and the differences are also a few kilometers per day, these differences are not only statistically significant, but also physically significant.
    The anomaly cosine correlation of ORAS5 with ADT for displacement speed is $0.54$ and RMSE is $2.22$ km/day.
    The anomaly cosine correlation of GFDL-OM4 is $0.55$ and RMSE is $1.49$ km/day.
    Accounting for lifetime when considering distance traveled improved the spatial pattern correlation, and resulted in a clearer, more statistically significant view of the differences.


\subsubsection{Ratio of Displacement to Distance Traveled} \label{sec:disp_dist}
    Eddies in GFDL-OM4 and ORAS5 live too long and travel too far, but have too little net displacement, meaning that eddies in GFDL-OM4 and ORAS5 do not move as linearly as eddies in ADT.
    In GFDL-OM4 both the mean ratio of displacement to lifetime and the mean ratio of distance to lifetime are deficient. 
    In ORAS5, the mean ratio of displacement to lifetime is deficient and the mean ratio of distance to lifetime is too large when compared to ADT.
    To further disentangle the relationships between lifetime, displacement, and distance traveled, we consider next the ratio of displacement to distance traveled, which is between 0 and 1.
    If the net displacement equals the total distance traveled then the ratio is 1 and the eddy has traveled in a straight line, whereas if the distance is much larger than the displacement then the ratio is small and the eddy has wandered or traveled in a circular path.
    
\begin{figure}[htbp]
        \centering
        \includegraphics[width=\linewidth]{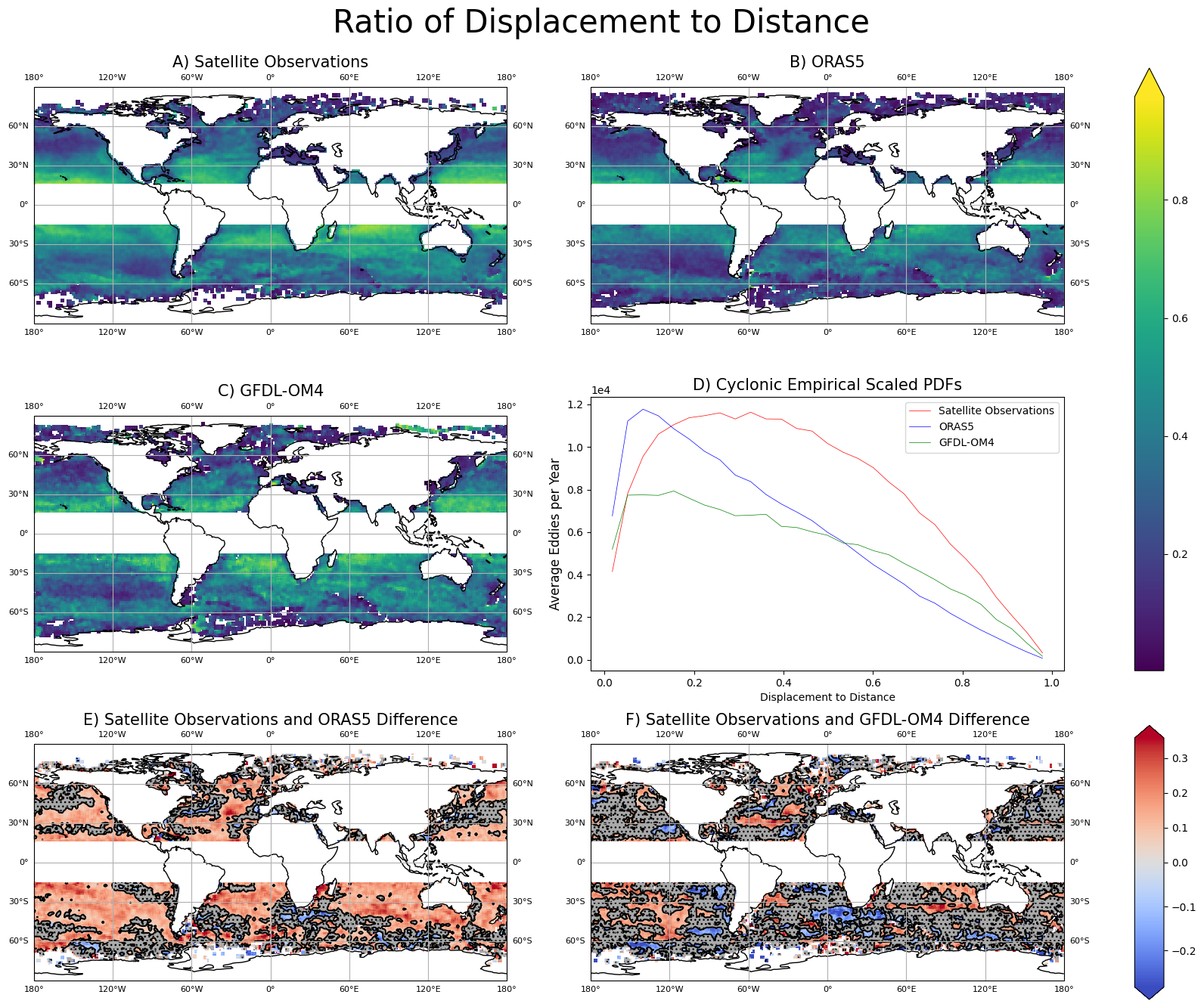}
        \caption{Ratio of Displacement to Distance. (A-C) display the spatial maps of the mean ratio of displacement to distance for ADT, ORAS5, and GFDL-OM4. (D) shows the scaled empirical global PDFs of the ratio of displacement to distance for ADT, ORAS5, and GFDL-OM4. (E-F) show the differences between ADT and ORAS5 maps and the differences between ADT and GFDL-OM4 maps for the ratio of displacement to distance. Areas that are grayed out do not have differences that are statistically significant.\label{fig:disp_dist}}
    \end{figure}

    The global scaled PDFs of the ratio of displacement to distance are displayed in Fig.\ \ref{fig:disp_dist}d. 
    GFDL-OM4 has a similar number of eddies to ADT with a very low displacement to distance ratios, i.e.\ eddies that wander rather than moving in a straight line.
    ORAS5, on the other hand, has to many eddies with a very low ratio of displacement to distance.
    Across the rest of the range of displacement to distance ratios both ORAS5 and GFDL-OM4 are missing a large number of eddies that propagate more linearly, with ratios greater than 0.5.

    Considering next the distribution of this ratio independent of the total number of trajectories, we find that the global cyclonic means of the ratio of displacement to distance traveled for ADT, ORAS5, and GFDL-OM4 are $0.41$, $0.33$, and $0.39$, the medians are  $0.39$, $0.29$, and $0.36$, and the standard deviations are $0.23$, $0.22$, and $0.24$, respectively. 
    Globally, eddies in ORAS5 and GFDL-OM4 do not propagate as linearly as the observed eddies in ADT do. ORAS5 is worse in this regard than GFDL-OM4.
    This is consistent with the fact that ORAS5 also is worse in distance traveled than GFDL-OM4 when comparing the global distributions to ADT. 

    The spatial mean maps are also displayed in Fig.\ \ref{fig:disp_dist}a-c. 
    Both ORAS5 and GFDL-OM4 perform relatively well at approximating the pattern observed in ADT well.  
    While in ORAS5 and GFDL-OM4, the ratio of displacement to distance is lower in the tropics, they accurately capture the broad spatial pattern. 
    Particularly both accurately represent the more linearly propagating eddies near the tropics. 
    The anomaly cosine correlation of ORAS5 to ADT is $0.79$ with a RMSE of $0.10$. The anomaly cosine correlation of GFDL-OM4 to ADT is $0.74$ with a RMSE of $0.09$. While ORAS5 is slightly better than GFDL-OM4 at locally approximating the pattern, the spatial pattern of eddy path straightness is incorrect in both data sets. 
    
\section{Summary and Discussion} \label{sec:conc}
    In this paper we compared mesoscale eddies identified by their SSH signature across three different datasets, all at eddy-permitting $\nicefrac{1}{4}^\circ$ resolution: observations (ADT), reanalysis data (ORAS5), and model output (GFDL-OM4).
    Across nearly all of the statistics presented, both ORAS5 and GFDL-OM4 are deficient.
    Not only are there not enough eddies in these datasets, but the resolved eddies tend to have about a 31\% weaker amplitude, have  about a 13\% larger radii, have 15\% more acircularity, and rotate about 19\% more slowly when compared to their observational counterparts. 
    Further, they tend to live about 12\% longer, move about 8\% less from birth to death and travel in the wrong direction, and their trajectories wander too much.
    Moreover, as eddy size is often overestimated in altimetry data \citep{bashmachnikov2020eddies}, the increased size of eddies in ORAS5 and OM4 suggests an even larger departure from reality.
    Data assimilation in ORAS5 seems to result in a modest improvement in accurately representing eddy area. 
    This suggests that there is some numerical smoothing in $\nicefrac{1}{4}^\circ$ ocean models that is stronger than the smoothing that occurs in the construction of the altimetry maps.
    
    Not only are these global attributes off, but the spatial distributions of many eddy characteristics are also off.
    Because of the tight coupling between mesoscale eddies and the large-scale circulation, many of these errors in eddy characteristics reflect biases in the large-scale circulation of the models.
    ORAS5 and GFDL-OM4 are able to accurately capture the spatial distributions of amplitude, equivalent radius, displacement, the ratio of displacement to lifetime, and the ratio of distance to lifetime. 
    For the other statistics, especially eddy lifetime and acircularity, ORAS5 and GFDL-OM4 generally perform poorly at representing the spatial distribution.
    For these other statistics, while ORAS5 and GFDL-OM4 may capture broad global trends, they are often missing many regions of important eddy behavior.
    Although one might expect eddy-permitting models to perform better at lower latitudes where the deformation radius is larger compared to the grid scale \citep{hallberg2013using}, our results do not show clear improvement in eddy representation at low latitudes. 
    However, it should be noted that we do not presents results in the middle equatorial band ($15^\circ$S-$15^\circ$N) due to concerns about the reliability of the altimetry maps in this region.
    Table \ref{tab:sim_table} collects all of the anomaly cosine similarities and RMSE for the figures in the previous sections as well as for their anticyclonic counterparts contained in the Supplementary Material.

    \begin{table}[ht]
    \centering
    \scriptsize
    \begin{longtable}{llrrrrrrrr}
    \toprule
     Section & Metric &  \multicolumn{4}{c}{Cyclonic} & \multicolumn{4}{c}{Anticyclonic }\\ 
     & & \multicolumn{2}{c}{ADT vs. ORAS5} & \multicolumn{2}{c}{ADT vs. GFDL-OM4} & \multicolumn{2}{c}{ADT vs. ORAS5} & \multicolumn{2}{c}{ADT vs. GFDL-OM4} \\
    \cmidrule(lr){3-4} \cmidrule(lr){5-6} \cmidrule(lr){7-8} \cmidrule(lr){9-10}
     &  & Sim & RMSE & Sim & RMSE & Sim & RMSE & Sim & RMSE \\
    \midrule
    \endfirsthead
    
    \toprule
     Section & Metric &  \multicolumn{4}{c}{Cyclonic} & \multicolumn{4}{c}{Anticyclonic }\\ 
     & & \multicolumn{2}{c}{ADT vs. ORAS5} & \multicolumn{2}{c}{ ADT vs. GFDL-OM4} & \multicolumn{2}{c}{ADT vs. ORAS5} & \multicolumn{2}{c}{ADT vs. GFDL-OM4} \\
    \cmidrule(lr){3-4} \cmidrule(lr){5-6} \cmidrule(lr){7-8} \cmidrule(lr){9-10}
     &  & Sim & RMSE & Sim & RMSE & Sim & RMSE & Sim & RMSE \\
    \midrule
    \endhead
    \ref{sec:census} & Eddy Census & 0.62 & 15.18 & 0.53 & 17.02 & 0.61 & 15.33 & 0.52 & 17.41 \\
    \ref{sec:amp} & Amplitude Mean (m) & 0.79 & 0.03 & 0.67 & 0.03 & 0.79 & 0.02 & 0.69 & 0.03 \\
    \ref{sec:area} & Equivalent Radius Mean (km)& 0.82 & 7.37 & 0.73 & 10.6 & 0.81 & 8.92 & 0.74 & 10.52 \\
    \ref{sec:shape} & Acircularity Mean & 0.33 & 4.83 & 0.22 & 6.45 & 0.35 & 4.83 & 0.21 & 5.56 \\
    \ref{sec:speed} & Speed Mean (m/s) & 0.78 & 0.05 & 0.68 & 0.07 & 0.80 & 0.04 & 0.75 & 0.06 \\
    \ref{sec:birth} & Track Birth & 0.58 & 0.17 & 0.47 & 0.19 & 0.57 & 0.16 & 0.46 & 0.18 \\
    \ref{sec:disp} & Disp Mean (km)& 0.85 & 61.21 & 0.75 & 76.99 & 0.72 & 80.74 & 0.51 & 120.96 \\
    \ref{sec:dist}  & Dist Mean (km) & 0.24 & 283.34 & 0.10 & 335.16 & 0.12 & 606.63 & 0.15 & 506.33 \\
    \ref{sec:life}  & Lifetime Mean (days) & 0.23 & 28.48 & 0.09 & 54.56 & 0.14 & 49.18 & 0.09 & 69.02 \\
    \ref{sec:disp} &  Disp to Life Mean (km/day) & 0.83 & 0.66 & 0.80 & 0.75 & 0.74 & 0.83 & 0.77 & 0.81 \\
    \ref{sec:dist} & Dist to Life Mean (km/day) & 0.54 & 2.22 & 0.55 & 1.49 & 0.51 & 2.17 & 0.54 & 1.63 \\
    \ref{sec:disp_dist}  &Disp to Dist Mean  & 0.79 & 0.10 & 0.74 & 0.09 & 0.72 & 0.10 & 0.66 & 0.09 \\
    \bottomrule
    \caption{Anomaly Correlation and Root Mean Square Error (RMSE) for Eddy Characteristics\label{tab:sim_table}}
    \end{longtable}
    \end{table}

    Eddy-permitting ocean models at $\sim 1/4^\circ$ resolution are the coarsest resolution models that permit the formation of mesoscale ocean eddies, so it is not surprising that their representation of eddies leaves something to be desired.
    Further, ORAS5 and GFDL-OM4 are only two representative datasets of reanalysis and model output. 
    It was not expected that either ORAS5 or GFDL-OM4 would exactly replicate the observations, yet the extent of ORAS5 and GFDL-OM4's deficiencies remains striking.
    One major difference between ORAS5 and GFDL-OM4 is that the former assimilates observational data while the latter does not.
    Although there are other differences between the models, it is likely that a large part of the improved performance of ORAS5 as compared to GFDL-OM4 in many metrics can be attributed to the assimilation of observational data.
    Nevertheless, the gap between the behavior of eddies in ORAS5 and in observations remains large.

    The reduced number, amplitude, and rotational speed of eddies in these eddy-permitting models, as well as their excessive width, might be attributable to the large viscosities used in these models.
    Large viscosities are needed in eddy-permitting models to help prevent spurious diapycnal mixing of tracers \citep{ilicak2012spurious,megann2021exploring}.
    These large viscosities dissipate excess kinetic energy \citep{pearson2017evaluation} which could account for there being too few and too weak eddies in the models evaluated here.
    
    Many recent backscatter parameterizations \citep{JHAH15,JDKO19,JAKK19,JDKOSSW20,Grooms23,yankovsky2024vertical} and machine-learning parameterizations \citep{ZB20,GZ21,PZAFZ24} are designed to improve the representation of eddies in eddy-permitting models.
    Neither of the models evaluated here use any of these recent backscatter or machine learning  parameterizations.
    These new parameterizations are often evaluated using metrics like kinetic energy, SSH, and SST variability, or temperature, salinity, and transport biases.
    The approach developed here promises to enable a  more fine distinction between the impact of these various parameterizations on the representation of eddies in eddy-permitting models.
    Eddy demographics could also potentially be used to guide the tuning of existing parameterizations, or the development of new ones.

    The westward propagation of eddies is driven by the interaction of the eddy with a wake of dispersive Rossby waves that it generates \citep{cushman1990westward,morrow2004divergent,early2011evolution}, while the random variability in eddy movement is driven by eddy-eddy interaction \citep{ni2020random}, and eastward propagation is primarily driven by large-scale currents or topographic interactions \citep{peng2021contrasting}.
    In the models evaluated here the eddies don't have realistic westward drifts; they instead tend to wander too much.
    This may reflect that eddy-eddy interactions are better represented in the models than the delicate interactions between eddies and Rossby wave that drive westward propagation.
    Indeed Rossby wave dynamics are known to be poorly represented near the grid scale at eddy permitting resolution \citep{wajsowicz1986free,dukowicz1995mesh}.
    Errors in the modeled propagation characteristics could also be driven in part by errors in eddy strength (amplitude, radius, and rotational speed), since eddy strength also affects propagation characteristics \citep{chen2022divergence}.
    If so, then backscatter parameterizations might improve eddy strength and therefore also eddy propagation.
    Similarly, the discrete representation of baroclinic instability at eddy-permitting resolutions is also known to be deficient \citep{barham2018some}; this could also contribute to eddies' being too weak and propagating incorrectly in eddy-permitting models.
    Improving the numerical discretizations used in the models might therefore improve the strength of modeled eddies, with potential consequent improvements in their propagation characteristics.

    There are many possible implications of this work. 
    Firstly, machine-learning models trained on $1/4^\circ$ degree data, either from GFDL-OM4 or ORAS5 will inherit the deficiencies of eddy representation uncovered by our work.
    Second, any scientific domain that is downstream of mesoscale eddies may be impacted due to the deficiencies uncovered. 
    \cite{cornec2021impact} shows that cyclonic eddies strengthen Deep Chlorophyll Maxima (DCM) which leads to favorable conditions for phytoplankton growth. 
    They conclude that eddy pumping is the dominant process for in strengthening the DCMs. 
    As the strength of eddy pumping is related to amplitude \citep{liu2023characteristics}, both ORAS5 and GFDL-OM4 may underestimate the strengthening of these DCMSs due to the globally deficient amplitude of these models.
    Likewise the lower speed and higher area of eddies in ORAS5 and GFDL-ORAS5 may also have implications on the presence of DCMs in these models, but the link between eddy area and speed with DCMs has not been thoroughly examined. 
    
    Future work will focus on examining high resolution, ``eddy-resolving" model output.
    There is some prior work in this area.
    \cite{ShiEtAl23} analyzed changes over time in the amplitudes of eddies in the META 3.2 data set.
    \cite{DingEtAl24} applied an in-house identification and tracking algorithm derived from \cite{chelton} to output from a simulation of the LICOM3 model at $\nicefrac{1}{10}^\circ$ resolution and compared the results to similar data derived from AVISO SSH observations.
    They found that the model, despite nominally having eddy-resolving resolution, had 30.8\% fewer eddies than observations; they also noted differences in the model-eddies' lifetime, amplitude, and radius for eddies that live longer than 28 days. They did not examine global or local differences in eddy speed average, acircularity, azimuth, displacement, distance traveled, or the ratio of displacement to distance.
    \cite{YHL24} applied the same eddy identification and tracking code that produces the META 3.2 dataset to output from ocean simulations at 10 km resolution using the POP model, and compared the results both to META 3.2 and across different scenarios. 
    They focused their analysis on eddy radius, lifetime, birth, and amplitude.
    They found fewer eddies in the model than in observations under present-day conditions, and predicted some changes in the amplitude, frequency, and radius of eddies in the future.
    These results suggest that even `eddy-resolving' models might benefit from backscatter parameterizations to increase their eddy census. 
    
    Further work in this area also suggests that ``eddy-resolving" models at $\nicefrac{1}{10}^\circ$ still may not have a high enough resolution to accurately resolve mesoscale features. 
    \cite{wang2020eddy} analyzes changes in eddy kinetic energy (EKE) in the Arctic Ocean when increasing model resolution from 4 to 1 km. 
    By increasing the resolution, they notice an increase of $40\%$ in EKE in the Arctic. 
    They conclude that in the Arctic, a 1 km resolution is needed to accurately resolve mesoscale dynamics.
    \cite{wekerle2020properties} compares eddies in two ocean models with a 1 km resolution in the Fram Strait and comes to a similar conclusion about the need for a 1 km resolution.

    Changes in eddy demographics from their preindustrial behavior (pre 1850) to the present and through the end of the century are also of interest for future investigation, yet cannot be estimated from the existing satellite record.
    If a high resolution model can be verified as accurately representing present-day mesoscale ocean eddy demographics, it could then be used to investigate both past and future eddies to assess the impacts of climate change on mesoscale ocean eddies.
    
    \section{Acknowledgements}
    This work was partially supported by NSF grants DMS-2310487 and OCE-2524611.

    The authors are grateful to F.\ Castruccio for discussions about the nature of ADT, SLA, and zos, and which variables are the most appropriate to compare.
    
    We acknowledge high-performance computing support from the Derecho system (doi:10.5065/qx9a-pg09) provided by the NSF National Center for Atmospheric Research (NCAR), sponsored by the National Science Foundation.

    This work utilized the Alpine high performance computing resource at the University of Colorado Boulder. Alpine is jointly funded by the University of Colorado Boulder, the University of Colorado Anschutz, Colorado State University, and the National Science Foundation (award 2201538). (https://doi.org/10.25811/k3w6-pk81)

    This work utilized the Blanca condo computing resource at the University of Colorado Boulder. Blanca is jointly funded by computing users and the University of Colorado Boulder. (https://doi.org/10.25811/v32c-gy42)

    We also thank the our editors Enrique Curchitser, Javier Arístegui, and Meng Xia as well as the six anonymous reviewers for their time and helpful comments that have substantially improved the quality of the manuscript.
    
\bibliography{biblio}

\section{Appendix A}

    \begin{table}[H]
        \footnotesize
        \centering
        \begin{tabular}{|l|lllllll|}
            \hline
             Section & Metric & Cyc Mean & Cyc Median & Cyc STD & Ac Mean & Ac Median & Ac STD \\ \hline
            \multirow{4}{*}{\ref{sec:amp}} & Amplitude ADT & 0.06 & 0.04 & 0.07 & 0.06 & 0.04 & 0.07 \\
            &Amplitude SLA & 0.06 & 0.04 & 0.08 & 0.06 & 0.04 & 0.06 \\
            &Amplitude Re & 0.04 & 0.02 & 0.06 & 0.04 & 0.02 & 0.06 \\
            &Amplitude OM4 & 0.04 & 0.02 & 0.04 & 0.05 & 0.02 & 0.07 \\
            \hline
            \multirow{4}{*}{\ref{sec:area}}&Equivalent Radius ADT & 50.71 & 46.14 & 20.61 & 52.38 & 47.17 & 22.49 \\
            &Equivalent Radius SLA & 50.71 & 46.07 & 20.43 & 52.04 & 46.99 & 21.69 \\
            &Equivalent Radius Re & 57.26 & 52.43 & 25.11 & 60.38 & 55.04 & 27.58 \\
            &Equivalent Radius OM4 & 58.81 & 54.85 & 25.31 & 59.84 & 55.32 & 26.04 \\ \hline
            \multirow{4}{*}{\ref{sec:shape}} &Acircularity ADT & 29.57 & 28.00 & 13.37 & 29.73 & 28.00 & 13.28 \\
            &Acircularity SLA & 29.17 & 27.50 & 13.24 & 29.54 & 28.00 & 13.16 \\
            &Acircularity Re & 34.19 & 33.00 & 14.19 & 34.68 & 33.50 & 14.22 \\
            &Acircularity OM4 & 36.97 & 36.00 & 14.47 & 34.60 & 33.50 & 14.75 \\\hline
            \multirow{4}{*}{\ref{sec:speed}}&Speed Average ADT & 0.16 & 0.12 & 0.12 & 0.15 & 0.12 & 0.10 \\
            &Speed Average SLA & 0.15 & 0.12 & 0.12 & 0.14 & 0.11 & 0.10 \\
            &Speed Average Re & 0.13 & 0.10 & 0.10 & 0.13 & 0.10 & 0.10 \\
            &Speed Average OM4 & 0.11 & 0.08 & 0.08 & 0.12 & 0.09 & 0.10 \\\hline
            \multirow{4}{*}{\ref{sec:life}} &Lifetime ADT & 93.51 & 71.00 & 69.20 & 96.42 & 72.00 & 75.42 \\
            &Lifetime SLA & 93.29 & 71.00 & 68.01 & 95.06 & 72.00 & 71.65 \\
            &Lifetime Re & 104.74 & 79.00 & 80.03 & 106.19 & 80.00 & 87.03 \\
            &Lifetime OM4 & 109.82 & 82.00 & 90.21 & 121.15 & 86.00 & 111.12 \\\hline
            \multirow{4}{*}{\ref{sec:disp}}&Displacement ADT & 206.37 & 132.77 & 242.49 & 203.47 & 134.51 & 240.31 \\
            &Displacement SLA & 218.31 & 145.23 & 243.93 & 214.86 & 146.94 & 239.04 \\
            &Displacement Re & 190.05 & 126.23 & 206.56 & 188.41 & 123.18 & 222.47 \\
            &Displacement OM4 & 175.13 & 119.41 & 182.92 & 180.87 & 118.45 & 233.08 \\ \hline
            \multirow{4}{*}{\ref{sec:disp}}&Ratio of Displacement to Lifetime ADT & 2.24 & 1.76 & 1.80 & 2.21 & 1.74 & 1.80 \\
            &Ratio of Displacement to Lifetime SLA & 2.38 & 1.91 & 1.81 & 2.37 & 1.90 & 1.80 \\
            &Ratio of Displacement to Lifetime Re & 1.98 & 1.50 & 1.72 & 1.96 & 1.45 & 1.88 \\
            &Ratio of Displacement to Lifetime OM4 & 1.77 & 1.37 & 1.49 & 1.68 & 1.28 & 1.47 \\ \hline
            \multirow{4}{*}{\ref{sec:dist}}&Distance ADT & 487.38 & 379.31 & 377.68 & 495.79 & 383.31 & 397.14 \\
            &Distance SLA & 490.48 & 383.92 & 376.88 & 499.65 & 391.99 & 387.88 \\
            &Distance Re & 667.46 & 488.89 & 624.42 & 668.45 & 479.27 & 816.43 \\
            &Distance OM4 & 509.02 & 379.05 & 512.70 & 511.18 & 365.72 & 555.98 \\ \hline
            \multirow{4}{*}{\ref{sec:dist}}&Ratio of Distance to Lifetime ADT & 5.37 & 4.94 & 2.34 & 5.38 & 4.85 & 2.53 \\
            &Ratio of Distance to Lifetime SLA & 5.43 & 5.00 & 2.36 & 5.51 & 5.01 & 2.52 \\
            &Ratio of Distance to Lifetime Re & 6.50 & 5.63 & 3.63 & 6.39 & 5.47 & 3.68 \\
            &Ratio of Distance to Lifetime OM4 & 4.87 & 4.15 & 2.82 & 4.49 & 3.79 & 2.66 \\ \hline
            \multirow{4}{*}{\ref{sec:disp_dist}}&Ratio of Displacement to Distance ADT & 0.41 & 0.39 & 0.23 & 0.40 & 0.39 & 0.23 \\
            &Ratio of Displacement to Distance SLA & 0.43 & 0.42 & 0.22 & 0.43 & 0.41 & 0.22 \\
            &Ratio of Displacement to Distance Re & 0.33 & 0.29 & 0.22 & 0.32 & 0.28 & 0.22 \\
            &Ratio of Displacement to Distance OM4 & 0.39 & 0.36 & 0.24 & 0.39 & 0.37 & 0.24 \\
            \bottomrule
            \end{tabular}
            \caption{Eddy Characteristic Summary Statistics. This table contains the mean, median, and standard deviation for both cyclonic and anticyclonic eddies for all the eddy attributes presented in the main paper.\label{tab:eddy_atts}}
    \end{table}

\end{document}